\documentclass[ras]{agutex}

\usepackage{graphicx}
\usepackage{amssymb}

\authorrunninghead{HELMBOLDT \& INTEMA}

\titlerunninghead{VLA Observations of Ionospheric/Plasmaspheric Waves}

\authoraddr{J.\ F.\ Helmboldt, US Naval Research Laboratory, 
Code 7213, 4555 Overlook Ave. SW, Washington, DC 20375 
(joe.helmboldt@nrl.navy.mil)}

\authoraddr{H.\ T.\ Intema, National Radio Astronomy Observatory, 
520 Edgemont Rd, Charlottesville, VA 22903 
(hintema@nrao.edu)}

\begin{document}

\title{Very Large Array Observations of Disturbed Ion Flow from the Plasmasphere to the Nighttime Ionosphere}

\authors{J. F. Helmboldt \altaffilmark{1} \& H. T. Intema \altaffilmark{2}}

\altaffiltext{1}{US Naval Research Laboratory, Washington, DC, USA.}
\altaffiltext{2}{Jansky Fellow of the National Radio Astronomy Observatory, Charlottesville, VA, USA.}

\begin{abstract}
We present the results of a multi-scale analysis of TEC fluctuations using a roughly five-hour observation of the bright radio source Virgo A with the Very Large Array (VLA) at 74 MHz in its B configuration.  Our analysis combines data sensitive to fine-scale structure ($\sim\! 10$ km and $<\!0.001$ TECU in amplitude) along the line of sight to Virgo A as well as larger structures (hundreds of km) observed using several ($\sim\! 30$) moderately bright sources in the field of view.  The observations span a time period from midnight to dawn local time during 1 March 2001. Several groups of magnetic eastward directed (MED), wavelike disturbances were identified and determined to be located within the plasmasphere ($2.1<L<2.9$).  We have also detected evidence of non-wavelike structures associated with these disturbances which are propagating roughly toward magnetic north.  These likely represent a non-uniform density flow from the plasmasphere toward the nighttime ionosphere.  AE and $K_p$ indices and GPS TEC data indicate that during the observations, there were low levels of geomagnetic activity accompanied by somewhat localized depletions in ionospheric density.  Thus, the observed plasmaspheric disturbance may be part of a flow triggered by these ionospheric depletions, likely associated with forcing from the lower atmosphere which is typically more prominent during quiet geomagnetic conditions.  In addition, we have also observed several roughly westward directed and southeast directed waves located within the ionosphere.  They are coincident in time with the plasmaspheric disturbances and may be related to the deposition of material onto the nighttime ionosphere.
\end{abstract}

\begin{article}

\section{Introduction}
Because of the negative impact on satellite-based communication, satellite operation, and GPS precision, disturbances within both the ionosphere and plasmasphere as well as interactions between the two have generated much interest and investigations.  These disturbances run the gamut from geomagnetic storms, substorms, streamers, ionospheric scintillations, storm enhanced densities, and traveling ionospheric disturbances (TIDs), to name a few.  Consequently, detailed explorations of these phenomena have been conducted using a wide variety of instruments including satellites, magnetometers, ionosondes, radar arrays, and GPS and other satellite beacon receivers.\par
Within this field of study, radio-frequency interferometers are a relatively underused resource.  Designed to operate as interferometric telescopes, such arrays are affected by the ionosphere and plasmasphere in the same way as ground-base satellite receivers.  Consequently, much effort has been devoted to compensating for these effects, especially within the VHF regime where the impact is much greater.  These corrections contain information about density fluctuations within the ionosphere and/or plasmasphere, being sensitive to the gradient of the total electron content (TEC) rather than the TEC itself \citep[see, e.g.,][]{jac92a,coh09,hel12a}.  When operating within the VHF regime and observing a bright source, interferometers can measure differences in TEC, or $\delta \mbox{TEC}$ between pairs of antennas to a precision of nearly $10^{-4}$ TECU (1 TECU$=10^{16}$ m$^{-2}$) and TEC gradients to a precision of $\sim\! 10^{-4}$ TECU km$^{-1}$ \citep[see, e.g.,][]{hel12a}.\par
Because of its relatively stable electronics, large collecting area, and available VHF system (bands at 74 and 330 MHz), the Very Large Array (VLA; located at $34^{\circ} \: 04' \: 43.497''$ N, $107^{\circ} \: 37' \: 05.819''$ W) has been used on a somewhat limited basis to study the ionosphere and plasmasphere.  Seminal investigations by \citet{jac92a} and \citet{jac92b} used 330 MHz observations with the VLA to identify and explore several wave-like phenomena identified within the data.  Follow-up work done by \citet{hoo97} showed that a new class of waves found by \citet{jac92a} were located within the plasmasphere.  They were shown to essentially be density fluctuations formed perpendicular to magnetic field lines at McIlwain $L$-parameters between about 1.8 and 4 which co-rotated with the plasmasphere with a moderate amount of (mostly) westward convection.  In addition, a statistical description of the typical, predominantly turbulent behavior of the ionosphere over the VLA using a large 74 MHz survey \citep{coh07} was presented by \citet{coh09}.  Recently, \citet{hel12b} detailed new spectral analysis techniques that enabled the simultaneous identification and characterization of wave-like disturbances within VLA VHF data from scales of a few kilometers up to more than 100 km.\par
Here, we detail a new interferometer-based approach that builds upon this previous work of \citet{jac92a}, \citet{hoo97}, and \citet{hel12b} by using a $\sim\!5$ hour 74 MHz observation of a very bright VHF source as well as several moderately bright sources spread over a $15^{\circ}$ area on the sky (or, about 80 km at an altitude of 300 km).  This new technique allows for a more thorough evaluation of the environment of fluctuations on medium to large scales than can be obtained with observations of a single source.  This multi-source method also has better precision than would be possible with a similar GPS-based approach.  When combined with simultaneous data for a single bright source in the field of view, we demonstrate that a unique characterization of both plasmaspheric and ionospheric wave properties from scales of a few to hundreds of kilometers can be attained.  In \S 2, we describe the observations and the processing/analysis of the data with more detailed descriptions included as Appendices.  In \S 3, we detail the properties of plasmaspheric disturbances within the data.  In \S 4, we similarly describe ionospheric disturbances contained within the data.  We include additional contemporaneous AE index and GPS data while interpreting the results in \S 5 and summarize in \S 6.

\section{Data and Analysis}
\subsection{Observations and Initial Calibration}
The data-set we have chosen to analyze (VLA program number AD441) is a roughly five hour long observation of the bright source Virgo A, or ``Vir A'' (also commonly known as 3C274) with the VLA  at 74 MHz.  Vir A is among the brightest cosmic VHF sources with a flux density at 74 MHz of about 1,700 Jy (1 Jy = $10^{-26}$ W m$^{-2}$ Hz$^{-1}$). The observations were taken in the post-midnight hours local time of 1 March 2001 starting at 06:08 UT using the VLA ``4-band'' system \citep{kas07}.  While observations with this system are generically referred to as 74 MHz observations, the central frequency was more precisely 73.8 MHz with a 1.5 MHz wide bandpass.  In this case, dual polarization was also used.  The $K_p$ index over this time period was 1--2 and the solar activity was fairly average ($F10.7\!\approx\!130$ SFU).  The VLA was in its B configuration with the antennas spanning an area about 11 km in diameter and laid out according to the diagram displayed in Fig.\ \ref{layout}.
\subsection{Spectral Analysis}
\subsubsection{Self-calibration Data}
The calibration, processing, and spectral analysis of the observations are described in detail in Appendices A--B.  Briefly, we used techniques developed by \citet{hel12a} to convert the observations of Vir A to measurements of the TEC gradient time series for each VLA antenna.  This conversion relies on a technique known as ``self-calibration'' which is discussed in detail in \citet{cor99}.  Hereafter, use of the term self-calibration in reference to any results presented implies that they were based on these measurements.  Two Fourier-based approachs were used to convert the TEC gradient time series (temporal sampling of 6.59 seconds) into mean spatial frequency spectra to identify disturbances that may be present within the data.  The first technique is described in detail by \citet{hel12b} which uses the phases of the temporal Fourier-transforms to determine a spatial frequency, $\xi$, and direction for each temporal mode.  Because it is based solely on phase information, we refer to this technique as ``phase-based.''  Within the phase-based technique, the determined direction of each mode is that of the line along which the phases of the Fourier transforms of the two TEC gradient components (i.e., north-south and east-west) agree.  This gives this approach the ability to sense the direction of motion for non-wavelike disturbances (see Appendix B for a more detailed discussion).\par
The second method uses a straightforward discrete Fourier transform (DFT) of the antenna-based temporal Fourier transforms to map the spectral power as a function of north-south and east-west spatial frequency.  This ``DFT-based'' method also employs an algorithm designed to mitigate the effects of the rather poorly behaved impulse response of the Y-shaped VLA within this procedure (again, see Appendix B for more discussion).  Because of the low spectral resolution of the relatively compact B-configuration VLA (11-km across), this technique still only allows one to reliably estimate a single spatial frequency and direction for each temporal mode (see Appendix B).  Since it is a more standard spectral approach, the DFT-based technique is mostly sensitive to wavelike disturbances for which it measures wavefront orientations and not necessarily directions of motion.  This makes the two techniques complementary when both wavelike and non-wavelike disturbances are present.\par
We have applied both techniques to the Vir A self-calibration data to produce two-dimensional spectral maps averaged within one-half-hour windows as functions of spatial frequency, $\xi$, and azimuth angle measured clockwise from north.  To aid with the identification of small-scale, small amplitude disturbances, we have normalized these spectra by dividing each $\xi$ bin by the total power over all azimuth angles within that bin.  For convenience, we will refer to these specially normalized spectra as ``azimuth spectra.''\par
The azimuth spectra derived from the Vir A data are displayed in Fig.\ \ref{paimg}.  The results for the phase-based technique are shown as greyscale images and the DFT-based results are plotted as color-coded contours.  The DFT-based azimuth spectra show several prominent features directed roughly toward magnetic east (azimuth of $102^\circ$ for the VLA) between local times of 01:30 and 03:00.  Their orientation implies that these disturbances are aligned along magnetic field lines.  They also have relatively large speeds, mostly spanning the range 200--400 m s$^{-1}$.  These properties imply that they are likely located within the plasmasphere and are the same magnetic eastward directed (MED) disturbances discovered by \citet{jac92b} with the VLA.  These were described in more detail by \citet{hoo97} as relatively long plasmaspheric irregularities stretched along field lines.  After 03:00, westward and southwest directed waves begin to dominate the DFT-based azimuth spectra.\par
The phase-based azimuth spectra also show detections of the same phenomena, but with notably different characteristics.  Specifically, the MED waves are broken into several groups and the estimated azimuths for each group seem to ``drift'' along linear structures toward the north with decreasing spatial frequency.  This may indicate that the MED waves are associated with a significant amount of non-wavelike structure which is moving in a different direction, possibly along the field lines toward magnetic north.  We will discuss these results and the properties of the MED waves in more detail in \S 3.  In contrast, the phase-based and DFT-based spectra agree rather well for the westward and southwestward directed phenomena seen after 03:00 local time, indicating they are likely dominated by wavelike structures.
\subsubsection{Field-based Calibration Data}
In addition to Vir A, several other moderately bright sources were also visible with the $15^\circ$-wide 74 MHz VLA field of view.  In particular, we identified 29 sources from the VLA Low Frequency Sky Survey \citep{coh07} which could be used to measure the TEC gradient as a function of time along the line of sight to each source, using a technique called ``field-based calibration'' \citep[see][and appendix A]{cot04}.  We will use the term field-based calibration to refer to any results that were based on this analysis.  Since the sources used within this technique were not nearly as bright as Vir A (between 2.5 and 10 Jy versus 1,700 Jy for Vir A), each of them could not be used to perform self-calibration.  This is because the relative contributions of other sources to the measured visibilities is significant for fainter sources.  For a source as bright as Vir A, these contributions are negligible.  However, for a moderately bright source, one can measure the effect of the ionosphere along the line of sight by making images of the source on relatively short time scales and measuring how its observed position changes with time \citep[see Appendix A and][for more details]{cot04}.\par
These position shifts yield measurements of the mean TEC gradient over the array toward a particular source, implying that these gradients have been smoothed with a roughly 11-km wide circular kernel (i.e., the diameter of the VLA B configuration).  In order to have a high enough signal-to-noise ratio within each image to detect the source, each image used one minute of data, implying that the data were effectively smoothed in time as well.  This spatial and temporal smoothing reduced the sensitivity of these measurements to small-scale, small amplitude fluctuations.  Specifically, the measured spectral power drops to one half the actual power at a spatial frequency of 0.04 km$^{-1}$ (or, a wavelength of 25 km) and/or at a temporal frequency of 26.5 hr$^{-1}$ (or, a period of 2.26 minutes).  The spectral response drops to zero at a wavelength of 11 km and/or a period of 1 minute.\par
When projected onto ionospheric heights (here, we used 300 km; see Appendix A), the pierce-points for these sources spanned nearly 100 km.  This is illustrated in Fig.\ \ref{calib} where we have plotted the projection of the VLA antennas onto a height of 300 km centered on each of the 29 sources.  With such a distribution of pierce-points, one can perform a straightforward spectral analysis by applying a Fourier transform in three-dimensions, one temporal and two spatial.  While such power spectrum ``cubes'' will be less sensitive to small-scale structures, they can be used to separate different Fourier modes with better fidelity than either the phase-based or DFT-based techniques applies to the self-calibration data.  This is because both of these methods rely on the conversion of temporal modes into spatial ones to enhance spectral resolution, and if there are several waves with the same temporal frequency but different spatial scales, only the strongest of the waves will be detected.  Thus, these two methods are quite complementary.\par
We have developed a procedure for using the 29 moderately bright sources to produce power spectrum cubes which is detailed in Appendix B.  We used this method to yield spectral cubes within the same half-hour intervals as were used for the azimuth spectra displayed in Fig.\ \ref{paimg} for the self-calibration data.  To display the wealth of information contained within these cubes, we have binned them by temporal frequency into four ``channels,'' with frequency ranges of 
0--6, 6--13, 13--21, and 21--30 hr$^{-1}$.  We have plotted the mean power spectra within each half-hour interval for each channel in Fig.\ \ref{chmap1}--\ref{chmap4} as greyscale contours.  For each of these ``channel maps,'' we have plotted the shape of the impulse response in the lower left corner as an ellipse that represents the full width at half maximum (FWHM) of the impulse response function (IRF).  We have also displayed maps of spectral power derived from the DFT-based technique applied to the self-calibration data as false-color images with the channel maps.  Each channel map shows varying levels of significant (i.e., $\gtrsim$ the size of the IRF) structure that, in some cases, overlap with features seen within the self-calibration-based maps.  These structures will be discussed further in \S 4.

\section{Plasmaspheric Disturbances}
As demonstrated in \S 2.2.2, our analysis of the Vir A-based self-calibration data has revealed a group of MED waves similar to those detected by \citet{jac92b} and established by \citet{hoo97} to be plasmaspheric irregularities.  We show the properties of these MED waves in more detail in Fig.\ \ref{med}.  For each of the three half-hour intervals between 01:30 and 03:00 local time, we have plotted in the left panels the spectral power versus spatial frequency projected along magnetic east, $\xi_{ME}$, for temporal modes with azimuths between $90^\circ$ and $180^\circ$ and speeds $<\! 2000$ km hr$^{-1}$ (556 m s$^{-1}$).  The latter criterion was used to exclude modes which were essentially noise.  Most of the MED waves have wavelengths between 25 and 100 km with a strong concentration near roughly 30 km.  This is very consistent with the range in wavelengths measured for similar plasmaspheric structures by \citet{jac96}.  For the first two half-hour intervals, there also appears to be a distinct group of relatively large waves (wavelength $\sim\! 140$ km) which have spectral power nearly an order of magnitude stronger than the bulk of the MED waves.\par
For the same temporal modes, we have also plotted spectral power as a function of speed projected along magnetic east, $v_{ME}$, in the middle panels of Fig.\ \ref{med}.  As noted before, most of the speeds are relatively high, ranging from about 150--500 m s$^{-1}$.  This is entirely consistent with the trace speeds observed for MED waves detected previously with the VLA \citep{hoo97}.  Assuming that the vast majority of the measured speed of any temporal mode is simply the projection of the co-rotation speed of the plasmasphere along the vector normal to the wavefronts (i.e., magnetic east), we can estimate the heights and L-shells of these detected MED waves.  The heights and L-shells corresponding to observed speeds of 100, 300, and 500 m s$^{-1}$ are labeled in the middle panels of Fig.\ \ref{med}.  They show that most of the MED waves are consistent with plasmaspheric heights between 2000 and 7000 km with a prominent peak at about 3300 km for the middle time interval.  For the geomagnetic latitudes of the VLA line of sight at these heights and times, this peak height corresponds to L$=2.27$.  The median L-shell for the VLA-detected MED waves detailed in \citet{hoo97} was 2.2, quite consistent with this estimate.  However, we note that the separate group of large/strong waves seen within the first two time intervals have speeds $<\!80$ m s$^{-1}$.  These are therefore likely ionospheric disturbances whose exact heights are less certain due to the influence of other factors on their motion such as the wind within the $F$-region.\par
As stated in \S 2.2.2, our data are entirely consistent with the model of \citet{jac96} and \citet{hoo97} for MED waves that describes them as plasmaspheric irregularities stretched out along field lines.  To estimate the zonal angular sizes of these disturbances as they map to lower altitudes, we have used the estimated heights and spatial frequencies to compute angular spatial frequencies, $\xi_{ang}$, in the magnetic eastward direction.  The power as a function of $\xi_{ang}$ is plotted in the right panels of Fig.\ \ref{med}.  From these plots, one can see that most of the structures occupy scales between about $0.1^\circ$ and $0.5^\circ$ with noticeable peaks in the second and third half-hour intervals corresponding to scales of about $0.3^\circ$ and $0.2^\circ$.\par
\begin{figure}
\noindent\includegraphics[width=\textwidth]{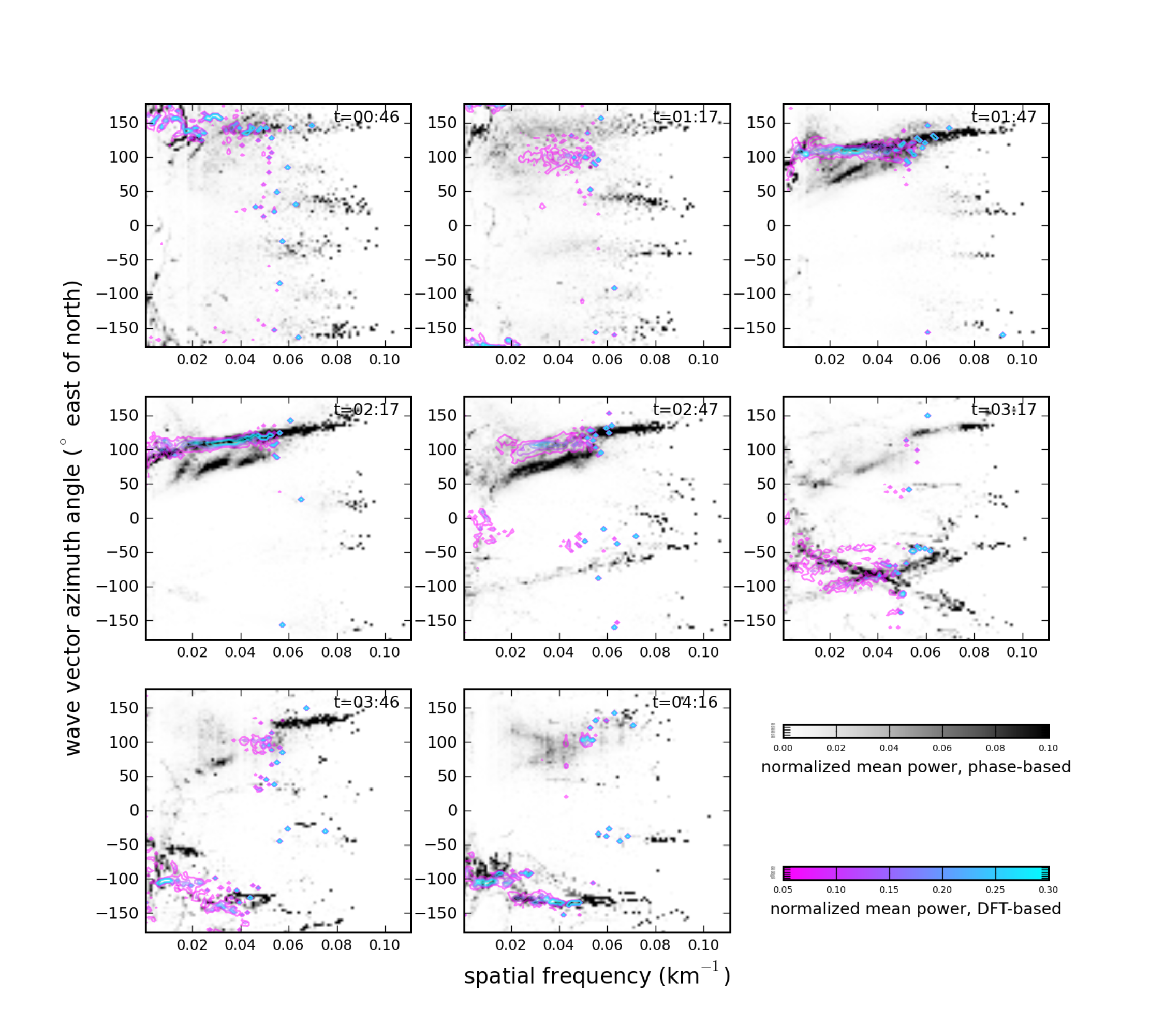}
\caption{For the Vir A self-calibration data, the mean power within bins of azimuth angle and spatial frequency with 1/2 hour bins for the phase-based (image) and DFT-based (contours) techniques (see \S 2 and Appendix B).  An azimuth of $0^\circ$ corresponds to north, $90^\circ$ to east, $-90^\circ$ to west, and $\pm180^\circ$ to south.  Within each spatial frequency bin, the values have been normalized by the total over all azimuth angle bins to make high-frequency features more apparent.  The local time at the center of each time bin is given in each panel.}
\label{paimg}
\end{figure}
\begin{figure}
\noindent\includegraphics[width=\textwidth]{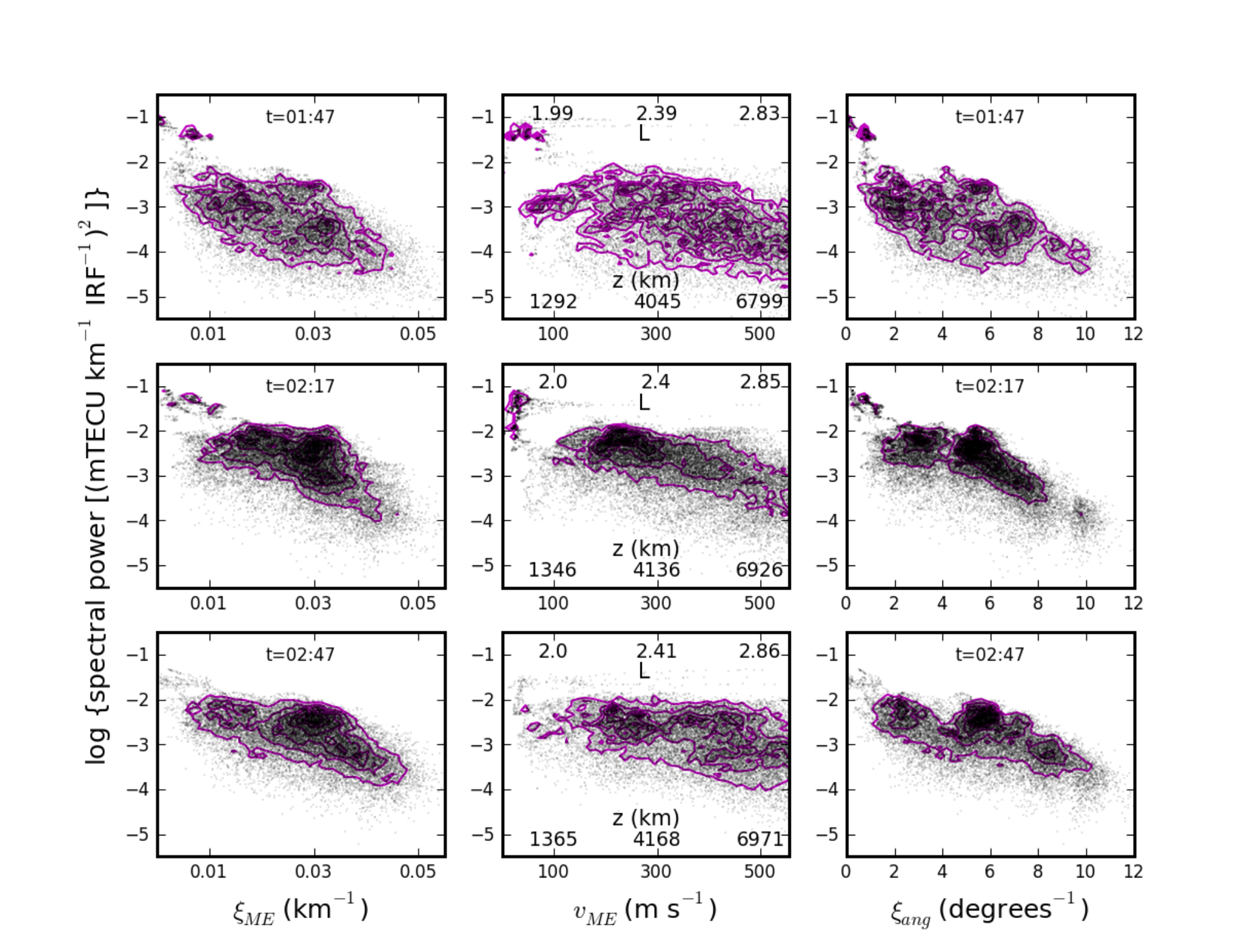}
\caption{For the three half-our time bins where they are most prominent (see Fig.\ \ref{paimg}), the properties of the field-aligned plasmaspheric disturbances.  The properties of each temporal mode from the DFT-based analysis with azimuths between $90^\circ$ and $180^\circ$ and speeds $<\!\! 2000$ km hr$^{-1}$ are plotted as separate points with magenta contours indicating the density of points within each panel.  The left panel shows the spectral power as a function of the spatial frequency vector projected along magnetic east, $\xi_{ME}$.  The middle panel plots the spectral power versus the velocity projected along magnetic east, $v_{ME}$, with corresponding estimates of the height, $z$, and L-shell computed assuming all of the motion is from co-rotation.  The right panels combine the results from the other two columns of plots to show the spectral power as a function of angular spectral frequency, $\xi_{ang}$ (i.e., the disturbance scale in geomagnetic longitude).}
\label{med}
\end{figure}

As noted in \S 2.2.2, the results from the phase-based spectral analysis technique show evidence for non-wavelike structure associated with the MED waves.  This is illustrated by the approximately linear features seen near magnetic east in Fig.\ \ref{paimg}.  We have postulated that this may indicate movement of non-wavelike structures northward of magnetic east.  Since these structures are likely associated with the plasmaspheric MED waves, the most logical direction of motion for such structures would be toward magnetic north along the field lines.\par
\begin{figure}
\noindent\includegraphics[width=\textwidth]{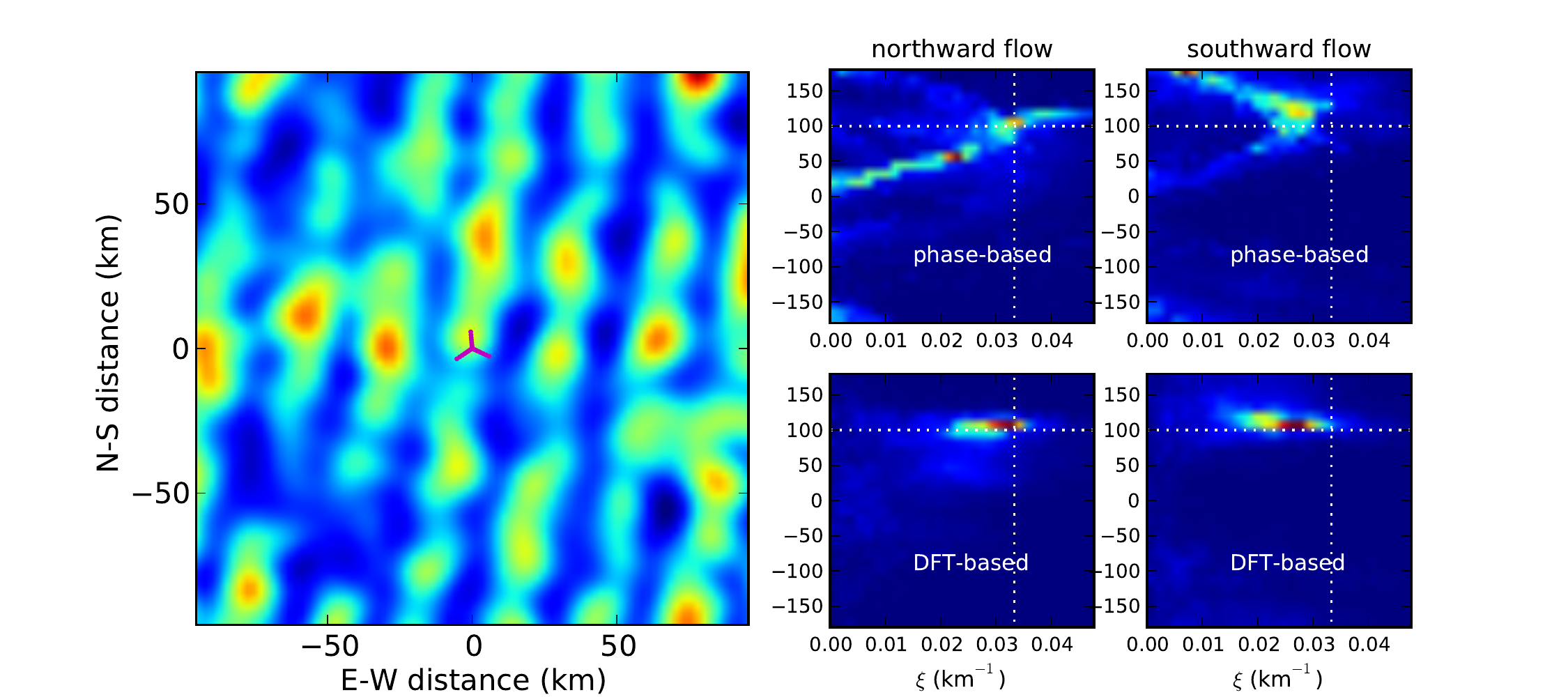}
\caption{Results from the ``toy'' model discussed in \S 3.  The left panel shows the initial distribution of $10^4$ Gaussian fluctuations modified by a 30-km-wavelength plane wave aligned along the magnetic field.  The positions of the VLA antennas are plotted as magenta points.  Results for both the phase-based and DFT-based analysis techniques are shown in the right panels for flows directed toward magnetic north (left panels) and toward magnetic south (right panels).  In all four panels, the spectral amplitude is displayed as a function of spatial frequency (abscissa) and azimuth (ordinate).  White dotted lines in all four panels indicate the actual azimuth and wavelength of the field-aligned plane wave.}
\label{toy}
\end{figure}

To demonstrate that this provides a good description of the phase-based results, we have constructed a ``toy'' model of these irregularities.  This model consisted of $10^4$ Gaussian TEC fluctuations with randomly generated sizes with a uniform distribution ranging from 15 to 40 km (here, ``size'' is $\sigma \sqrt{2\pi}$, where $\sigma$ is the standard deviation for the Gaussian).  The amplitudes of the Gaussians were drawn randomly from a Rayleigh distribution with a mode of unity (arbitrary units).  The amplitudes of these Gaussians were then modified by a cosine density perturbation with a wavelength of $\lambda = 30$ km and an azimuth of $100^\circ$ by multiplying by a factor of $[1+0.5\mbox{cos}(2\pi \vec{r}_G \cdot \hat{r}_w / \lambda)]$ where $\vec{r}_G$ is the position vector of each Gaussian and $\hat{r}_w$ is a unit vector perpendicular to the cosine wavefronts.  The initial field of fluctuations is displayed in the left panel of Fig.\ \ref{toy} with the positions of the VLA antennas plotted as points in the center for reference.\par
Within the model, both the Gaussian fluctuations and the cosine plane wave were given co-rotation speeds of 100 m s$^{-1}$ due east.  The Gaussian fluctuations were also given an additional flow velocity parallel to the cosine wavefronts of 200 m s$^{-1}$.  We then simulated VLA measurements by advancing the field of fluctuations using these constant velocities with 10-second time steps over a one-hour time period.  We ran the simulation twice, once with the flow velocity directed toward magnetic north, and once toward magnetic south.  As with the actual self-calibration data, we fit a second order two-dimensional polynomial to the simulated $\delta \mbox{TEC}$ measurements (see Appendix A) and applied both the phase-based and DFT-based spectral analysis methods to the results.  As with the real VLA data, we used a sliding half-hour wide boxcar, thus yielding dynamic spectra over the central half-hour of the simulated observations.\par
The results for the phase-based and DFT-based analyses for both the northward and southward flows are shown in the right panels of Fig.\ \ref{toy} as images of the spectral amplitude as a function of spatial frequency and azimuth.  As one would expect, the DFT-based approach successfully recovers the properties of the cosine wave with little evidence of the Gaussian fluctuations, albeit with a somewhat overestimated size for the southward flow (see the bottom right panel of Fig.\ \ref{toy}).  Conversely, the phase-based spectra show ample evidence of the influence of the Gaussian disturbances.  For both the northward and southward flows, the measured azimuths skew toward the direction of motion for the Gaussians toward lower frequencies where they tend to dominate the spectral power.  They form roughly linear structures very similar to those evident in Fig.\ \ref{paimg}, even reproducing the ``overshoot'' to more southeastward azimuths (i.e., $>\! 100^\circ$) for higher spatial frequencies seen within some of the features in Fig.\ \ref{paimg}.\par
The exact shapes of these spectral features depend on several factors such as the Gaussian fluctuations' distributions of sizes and amplitudes and their flow speed as well as the number, sizes, and strengths of the field-aligned cosine waves.  It is unclear to what degree the structures detected within our phase-based spectra constrain the properties of the flow(s) and associated MED waves.  However, it is clear that our toy model has demonstrated that a substantial flow of non-uniform density material along field lines toward magnetic north provides a adequate explanation of the structures observed within the phase-based azimuth spectra.  Given that the VLA lines of sight throughout the observations probed latitudes above the geomagnetic equator, this implies flow of material down flux tubes.  This is consistent with the fact that the observations were conducted at night when some refilling of the ionosphere from the plasmasphere is expected \citep[see, e.g.,][]{par74}.

\section{Ionospheric Waves}
\begin{figure}
\noindent\includegraphics[width=\textwidth]{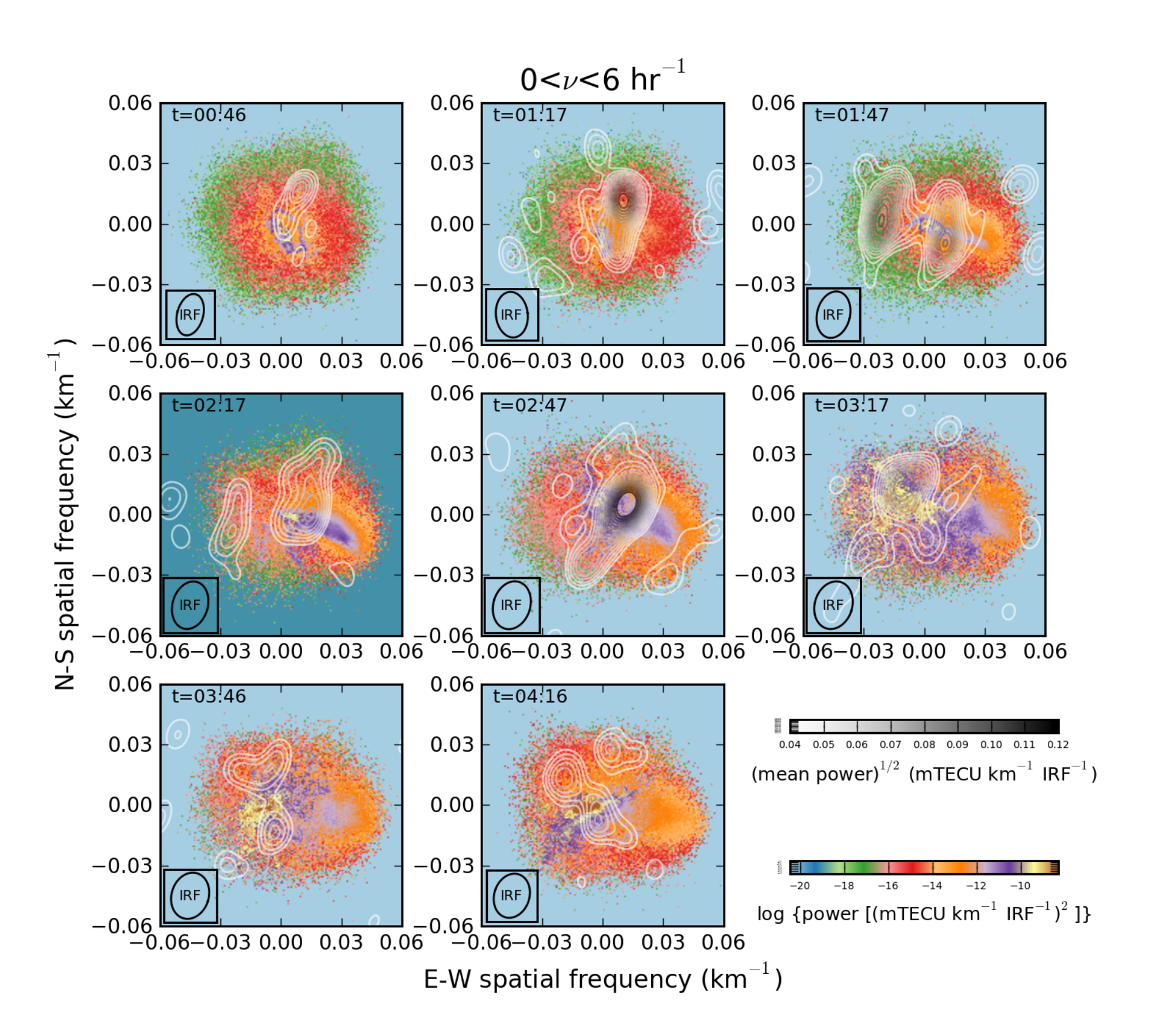}
\caption{Within 1/2 hour windows, the mean spectral power as a function of spatial frequency for the field-based calibration data (see \S 2 and Appendix B) for temporal frequencies between 0 and 6 hr$^{-1}$ plotted as greyscale contours.  The shape of the impulse response function (IRF) for each map is plotted in the lower left of each panel as an ellipse representing the full width at half maximum (FHWM) of the IRF.  The underlying image within each panel is a map of the log of the total power from the self-calibration data using the DFT-based spectral analysis technique (see \S 2, Appendix B, and Fig.\ \ref{paimg}).}
\label{chmap1}
\end{figure}

\begin{figure}
\noindent\includegraphics[width=\textwidth]{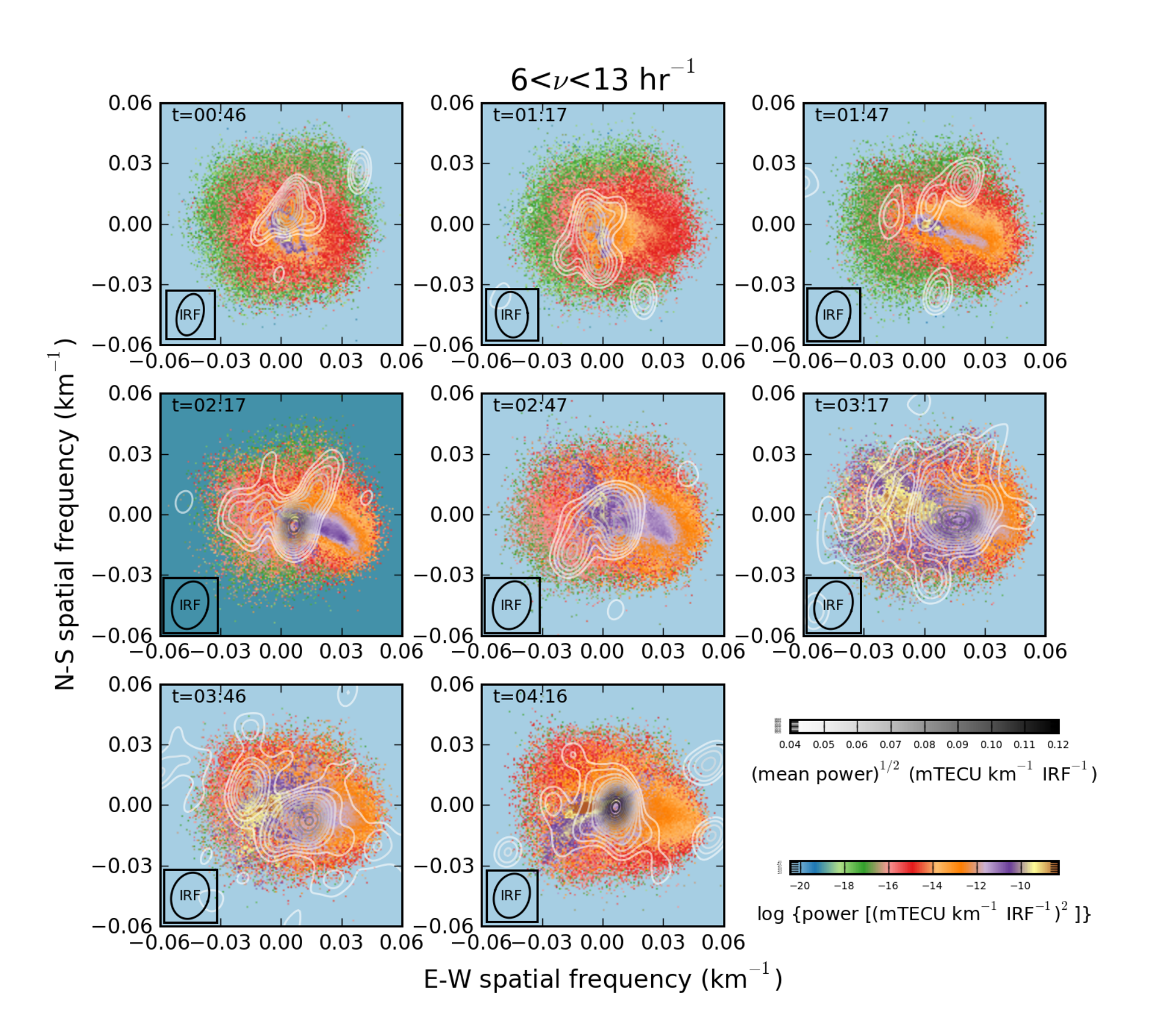}
\caption{The same as Fig.\ \ref{chmap1}, but for temporal frequencies between 6 and 13 hr$^{-1}$.}
\label{chmap2}
\end{figure}

\begin{figure}
\noindent\includegraphics[width=\textwidth]{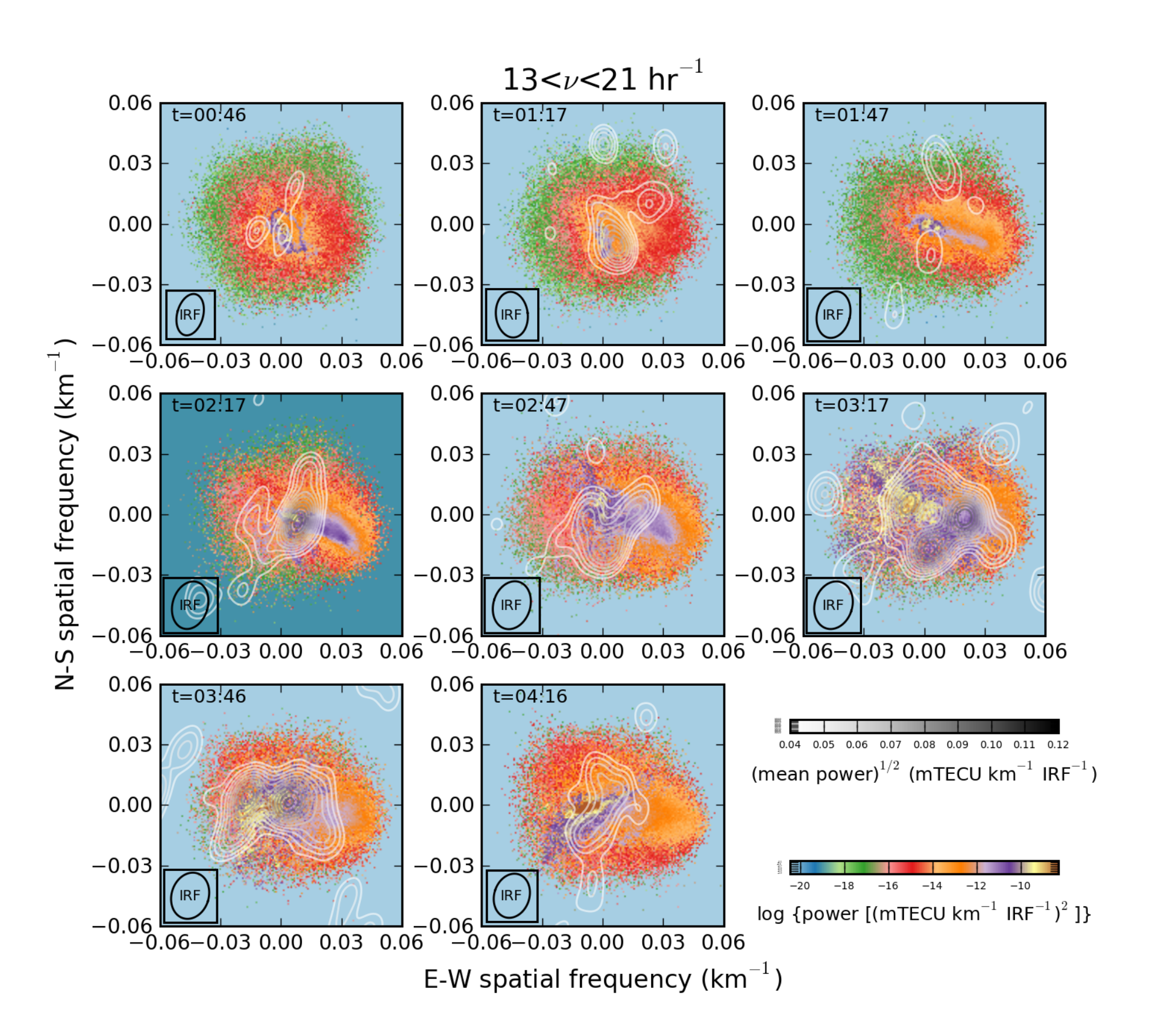}
\caption{The same as Fig.\ \ref{chmap1}, but for temporal frequencies between 13 and 21 hr$^{-1}$.}
\label{chmap3}
\end{figure}

\begin{figure}
\noindent\includegraphics[width=\textwidth]{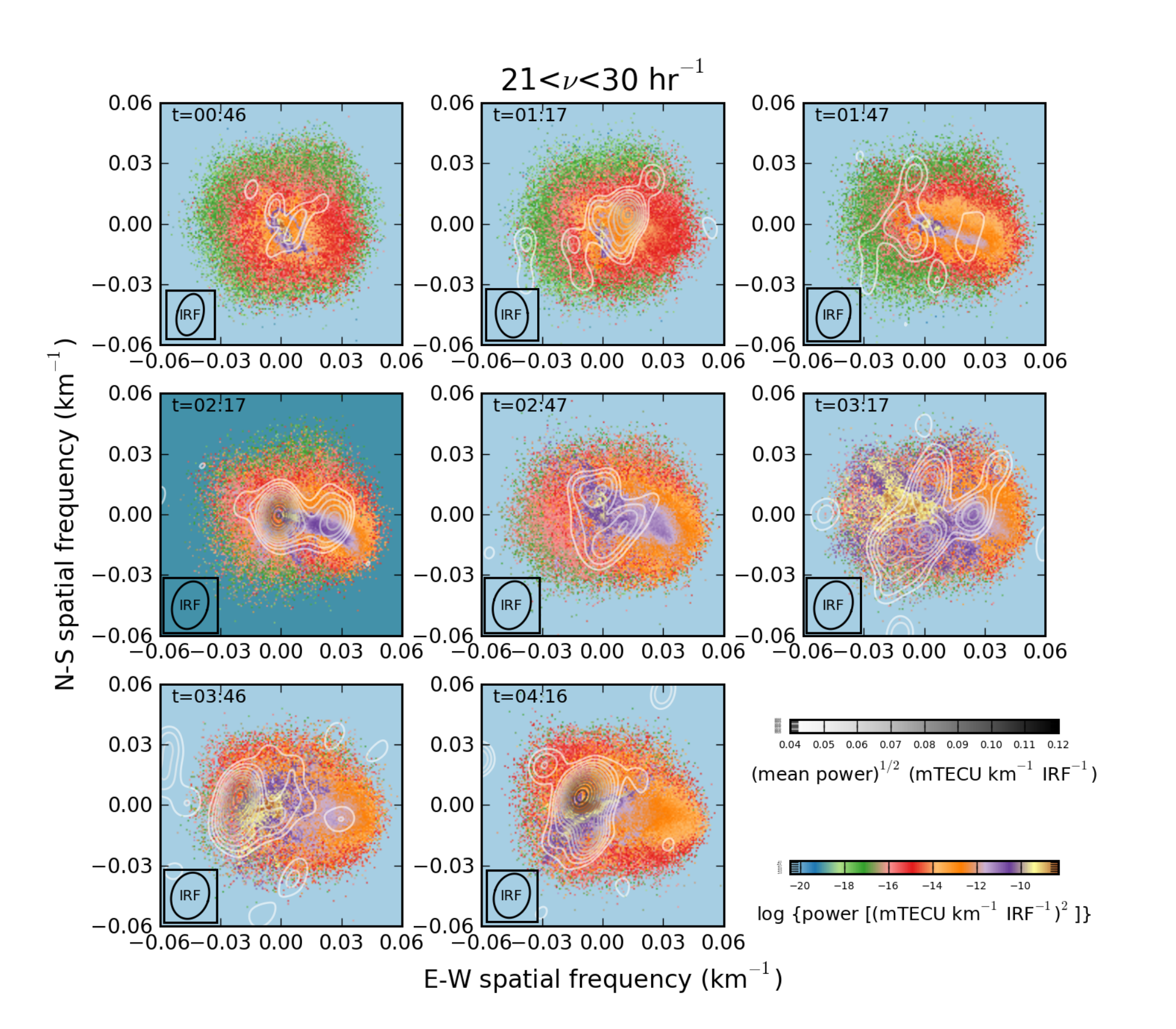}
\caption{The same as Fig.\ \ref{chmap1}, but for temporal frequencies between 21 and 30 hr$^{-1}$.}
\label{chmap4}
\end{figure}

In addition to the significant presence of plasmapheric activity within the data, there is also evidence of ionospheric disturbances, chiefly among the field-based calibration data.  Since the field-based data are based on observations of several sources with non-parallel lines of sight, it is much more effective at probing ionospheric rather than plasmaspheric structures.  For instance, at a height of 300 km, the mean separation among the lines of sight to the 29 sources used is about 12 km.  This implies an approximate Nyquist sampling limit of a minimum of 24-km-sized structures.  Indeed, the channel maps shown in Fig.\ \ref{chmap1}--\ref{chmap4} show little if any structure beyond spatial frequencies of $1/12=0.042$ km$^{-1}$.  If we increase the assumed height to 3000 km, the approximate height of the plasmaspheric MED waves described in the previous section, the minimum probable scale likewise increases to 240 km.  This is much too large to detect any of the plasmaspheric MED waves which have wavelengths $\lesssim \! 100$ km.\par
The channel maps in Fig.\ \ref{chmap1}--\ref{chmap4} give a glimpse of the large amount of information about ionospheric disturbances contained within the field-based spectral cubes.  At the beginning of the observing run, relatively low-frequency, northeastward-directed waves are prominent with weaker, higher-frequency waves directed toward the south/southeast.  These all have wavelengths between 50 and 100 km.  Starting at a local time of 01:47, the ionospheric MED waves described in \S 3 appear over a wide range of temporal frequencies, as high as 20 hr$^{-1}$.  Unlike the self-calibration data, they are seen within the field-based data through the remainder of the observing run.  Westward/southwestward directed waves are seen to overlap with those detected with the self-calibration data starting at a local time of about 03:30, most prominently for frequencies $>\!6$ hr$^{-1}$.  However, unlike the self-calibration data, roughly westward (some northwestward, some southwestward) waves are apparent at all times starting at the third time bin centered at a local time of 01:47.  They span nearly the entire range of temporal frequency and have wavelengths of between 50 and 100 km, similar to the ionospheric MED waves.

\section{Interpretation}
The relatively low $K_p$ indices for the time of our observations and the days preceding it ($\leq\!3$) indicate that the plasmaspheric disturbances we observed are not related in any way to geomagnetic storm activity.  They instead appear to be the same plasmaspheric irregularities elongated along magnetic field lines discovered by \citet{jac92b} and described in more detail by \citet{jac96} and \citet{hoo97}.  However, unlike the methods employed by these authors which rely on fitting a single plane wave to the VLA $\delta \mbox{TEC}$ measurements for a given time interval, our spectral analysis techniques can break the data up into many wavelike components.  These techniques reveal that at a given time, there can be many plasmaspheric MED waves occupying a wide range of L-shells and sizes as well as ionospheric MED waves.  In addition, our specialized phase-based spectral analysis method has revealed the existence of non-uniform density flows of material down the flux tubes occupied by these irregularities toward the nighttime ionosphere.\par
In order for such a flow to take place, there needs to be some depletion within the ionosphere.  Of course, during the night, recombination naturally causes such a depletion.  However, it is well established that during times of low geomagnetic activity, the density within the ionosphere can fluctuate substantially, between about 25\% and 40\% on scales of a few hours to several days \citep{for00,mik04,mik07}.  It may be that during our VLA observations, larger than average nighttime depletions occurred as a result of quiet-time variability caused by forcing from the lower atmosphere via planetary waves, gravity waves, and/or tides \citep{che92,ris01,liu10}.\par
\begin{figure}
\noindent\includegraphics[width=\textwidth]{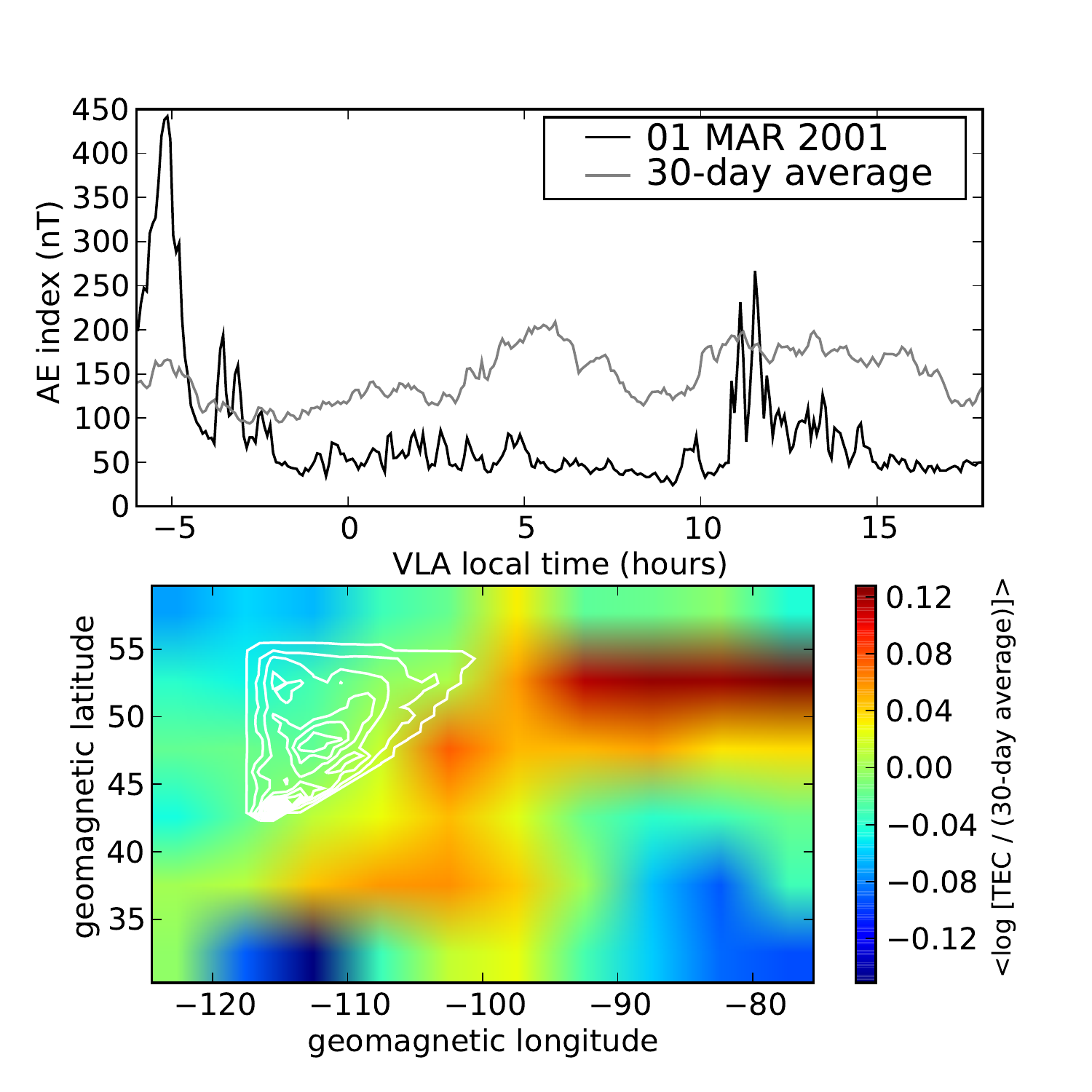}
\caption{Upper:  The AE index averaged within 5-minute bins for the date of the VLA observations, 1 March 2001 (black curve), and averaged with $\pm15$ days of the observations (grey curve) as functions of VLA local time.  Lower:  For the time range during the VLA observations ($06\!<\!$UT$\!\!<\!\!11$), the mean of the log of the ratio of the vertical TEC on 1 March 2001 to the 30-day average (centered on 1 March 2001) within $5^\circ\!\times\!5^\circ$ bins in geomagnetic latitude and longitude from the MIT Madrigal database.  White contours represent the spectral power of field-aligned plasmaspheric disturbances as a function of geomagnetic longitude and estimated invariant latitude, $\Lambda$.}
\label{tecrat}
\end{figure}

In the upper panel of Fig.\ \ref{tecrat} we establish that indeed, the level of geomagnetic activity was rather low during our observations.  We have done this by plotting the AE index measure within 5-minute intervals on the date of the observation, 1 March 2001, and the mean within $\pm \! 15$ days, both as functions of VLA local time.  While evidence of substorm activity can be seen near sunset and midmorning, the AE index during the time of our observations was substantially lower that the typical value for that time of year (i.e., late winter/early spring 2001).\par
To show that this low level of activity was accompanied by ionospheric depletions, we obtained GPS-based vertical TEC maps for North America from the MIT Haystack Observatory Madrigal Database for 1 March 2001 and for 15 days before and after this date.  We used these data to construct maps of vertical TEC at $5^\circ \! \times \! 5^\circ$ resolution for geomagnetic longitudes between $-125^\circ$ and $-75^\circ$ and geomagnetic latitudes from $30^\circ$ to $60^\circ$ at 5-minute intervals.  We then computed a 30-day mean map at each 5-minute interval and divided it into the corresponding map for 1 March 2001 to remove typical intraday variations.  We then computed the mean of the log of this ratio at each latitude and longitude during the time of our observations and have displayed the result in the lower panel of Fig.\ \ref{tecrat}.  From this map, one can see that there were several persistent, localized depletions relative to what was typical for the late winter/early spring of 2001.  There were also several locations of relatively enhanced TEC,
but most of the area in the map shown seems to be below average.  This is qualitatively consistent with the idea that these variations are the result of forcing from the lower atmosphere which can have a non-uniform effect on the observed TEC \citep[see, e.g.,][]{liu10}.\par
For comparison, we have also plotted contours of spectral power as functions of estimated geomagnetic longitude and invariant latitude, $\Lambda$, for the plasmaspheric MED waves described in \S 3 and Fig.\ \ref{med}.  This shows that the L-shells occupied by these irregularities map well to relatively depleted regions of the ionosphere.  It may then be that the irregularities themselves are tied to substructure within these depleted regions.  Such structure may cause some flux tubes to be more depleted than others as they refill the ionosphere, causing wavelike features as the ion density fluctuates as a function of geomagnetic longitude.  If there are in fact significant structures on scales of a few tenths of a degree in geomagnetic longitude (see \S 3 and the right panels of Fig.\ \ref{med}), this could contribute to the formation of the observed plasmaspheric MED waves.  The GPS-based data does not provide the level of spatial resolution required to search for such structures.  Our own VLA data has the ability to detect features on these scales, but if they exist, they are unfortunately overwhelmed by the plasmaspheric MED waves and ionospheric disturbances.  Future studies based on satellite beacon data may be able to better test this hypothesis.\par
There are two classes of ionospheric disturbances coincident in time with the plasmaspheric disturbances, field-aligned waves directed toward magnetic east and roughly westward directed waves.  Both classes of waves have wavelengths between about 50 and 100 km, and therefore can be categorized as medium scale traveling ionospheric disturbances (MSTIDs).  However, they occupy a relatively large range in temporal frequencies, including higher frequencies than have been previously observed.  With amplitudes $\lesssim \! 0.01$ TECU, they are also weaker than typical MSTIDs \citep[see, e.g.,][]{her06,tsu07}.  This is likely a reflection of the limitations in sensitivity and time resolution of the instruments used to characterize MSTIDs in past studies (typically GPS receivers).\par
The fact that they occur at the same time as the observed plasmaspheric disturbances may indicate that the generation of these waves may be related to the deposition of material from the plasmasphere onto the ionosphere.  The MED, field-aligned waves may be generated in the topside ionosphere where other factors such as winds will have less of an effect and the disturbances are more likely to remain aligned with the magnetic field.  The westward directed waves may arise lower within the $F$-region where the wind has more of an influence.  According to the publicly available GPI data-driven runs of the TIEGCM code \citep[][currently available at http://www.hao.ucar.edu/modeling/tgcm/]{wan99}, the zonal wind above about 350 km is generally westward near the location of the VLA during the time of day and year of our observations.
\section{Conclusions}
Using the unique capabilities of the VLA VHF system, we have developed a multi-scale approach to the spectral analysis of TEC gradients.  Probing a wide range of spatial scales and amplitudes, we have identified both plasmaspheric and ionospheric waves present during a $\sim\!5$ hour observation during the night of 1 March 2001.  Contemporaneous data suggest that geomagnetic activity was relatively low during this time, implying that the plasmaspheric disturbances were not part of storm-related activity.   Instead, combining the VLA data with contemporaneous AE index and GPS data, we have concluded that the field aligned waves in the plasmasphere appear to be associated with the flow of non-uniform density material along their wavefronts toward relatively depleted regions of the ionosphere.  The low level of geomagnetic activity and the spatial distribution of ionosphere depletions hint that the depletions may be linked to density fluctuations caused by forcing from the lower atmosphere via planetary waves, gravity waves, and/or tides.  We are currently conducting a more detailed investigation of this possibility using seven years of GPS-based TEC maps, AE index data, and 74 MHz VLA data from the VLA Low-frequency Sky Survey \citep{coh07}.\par
We have also simultaneously detected relatively fast (roughly 150--350 m s$^{-1}$) and weak (amplitudes $\lesssim \! 0.01$ TECU) MSTIDs within the ionosphere.  They were coincidence in time with the plasmaspheric disturbances and may have been generated by the deposition of material from the plasmasphere onto the nighttime ionosphere.  They were directed rough toward the west and toward magnetic east.  This may be a reflection of the altitudes at which these waves were generated.  The field-aligned waves may have been generated relatively high within the ionosphere whereas the westward moving waves may have formed lower in the middle $F$-region where the wind has a larger effect.
\appendix
\section{Data and Calibration}
\subsection{Initial Calibration}
The data were obtained from the National Radio Astronomy Observatory (NRAO) data archive\footnote{https://archive.nrao.edu} and initially processed within the Astronomical Image Processing System \citep[AIPS;][http://www.aips.nrao.edu]{bri94}.  Within AIPS, the complex bandpass response of each antenna was computed and the data were corrected for this.  An initial calibration was performed using a model of the distribution of the intensity of Vir A on the sky presented by \citet{kas07} (currently publicly available at http://lwa.nrl.navy.mil/tutorial/).  This model was used to compute model values of the correlations of signals between pairs of antennas, or ``visibilities.''  This model visibility data-set was then divided into the observed visibilities and the antenna-based gains were solved for within each scan (i.e., a contiguous block of observing time) using standard AIPS tasks.  This calibration technique is commonly referred to as ``self-calibration''; its application and robustness are detailed in \citep{cor99}.  In this instance, self-calibration was used over relatively long time intervals to correct for amplitude and phase variations in the responses of the individual antennas which generally vary on relatively long time scales.  This also has the effect of removing any slowly varying ionospheric contribution to the observed phases, but this has little impact on our analysis since we are focussing on variations on much shorter time scales (see below).
\subsection{Self-calibration}
Following the initial calibration, a second round of self-calibration was performed, this time at much shorter time intervals of 6.59 seconds, the shortest possible for this data-set given the time-averaging used to measure the visibilities.  The motivation for performing two separate rounds of self-calibration will become apparent in the following section where we will detail a separate calibration performed over the entire 74 MHz field of view.\par
\begin{figure}
\noindent\includegraphics[width=\textwidth]{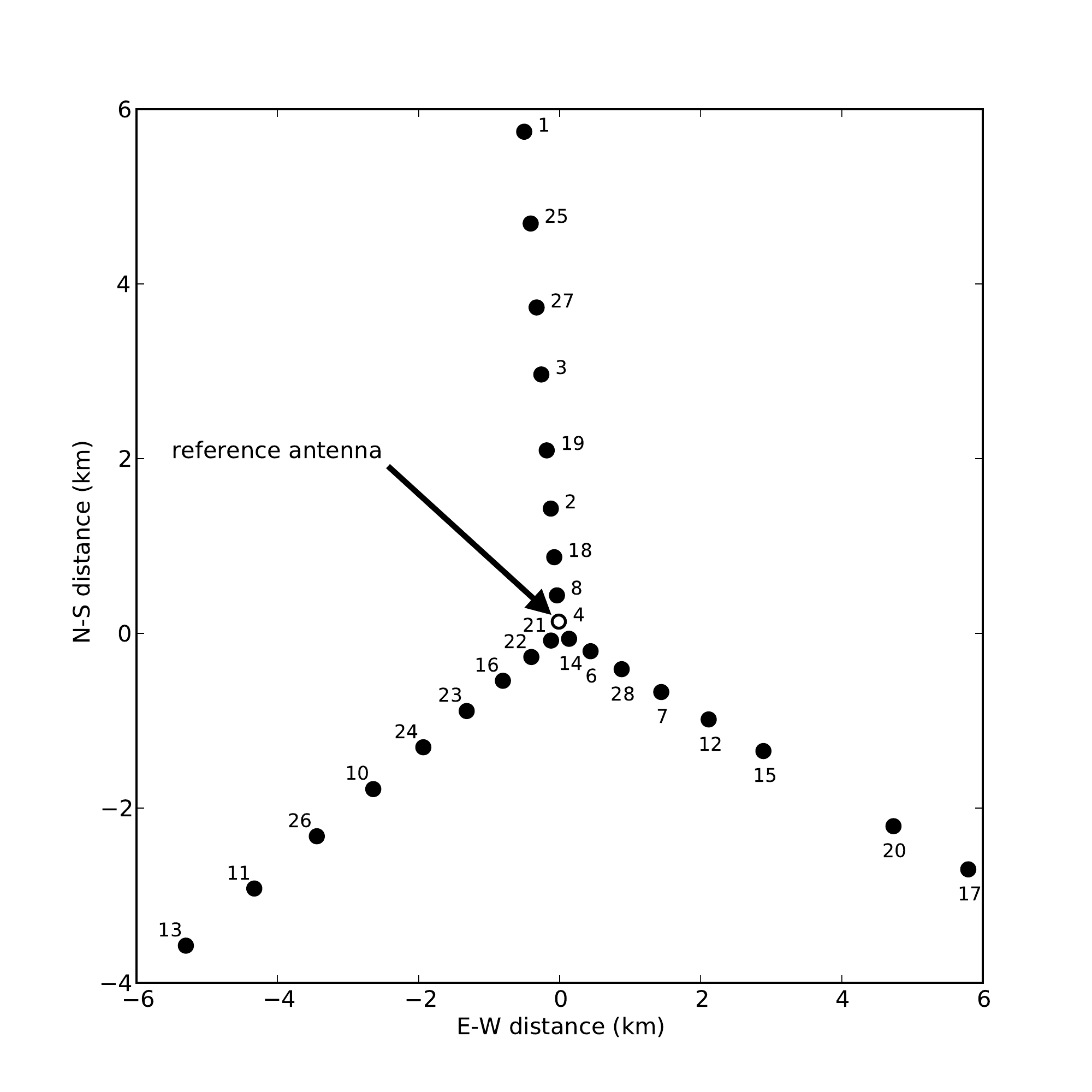}
\caption{The layout of the VLA antennas during the observations.  The reference 
antenna (see Appendix A) is highlighted in white.}
\label{layout}
\end{figure}

This second round of self-calibration was used to compute the difference in TEC between each antenna and the reference antenna \citep[see][and Fig.\ \ref{layout}]{cor99} according to \citet{hel12a}.  Unlike the data-set used by \citet{hel12a}, we found no obvious jumps in the self-calibration determined antenna phases which would negatively impact the process of unwrapping the phases needed to compute the values of $\delta \mbox{TEC}$ as functions of time.  Instead, we were able to simply unwrap the phases as functions of time within each scan and polarization.  The phases were then de-trended by subtracting smoothed versions of the phase time series using a 1/2 hour wide, sliding boxcar filter.  Near the edges of each scan, a linear fit to the phases as functions of time were also subtracted to mitigate edge effects within the smoothing process.  The boxcar was chosen to have the narrowest width that would not substantially dampen the longest period waves expected within the spectral analysis to be perform (see \S 4).  We expect to see waves with periods up to about one hour \citep[e.g.,][]{her06,hel12b}.  The subtraction of a version of the data smoothed with a 1/2 hour wide boxcar will reduce the amplitudes of such waves by a factor of about 2.75 which is acceptable given that these waves also tend to have relatively large amplitudes.\par
\begin{figure}
\noindent\includegraphics[width=\textwidth]{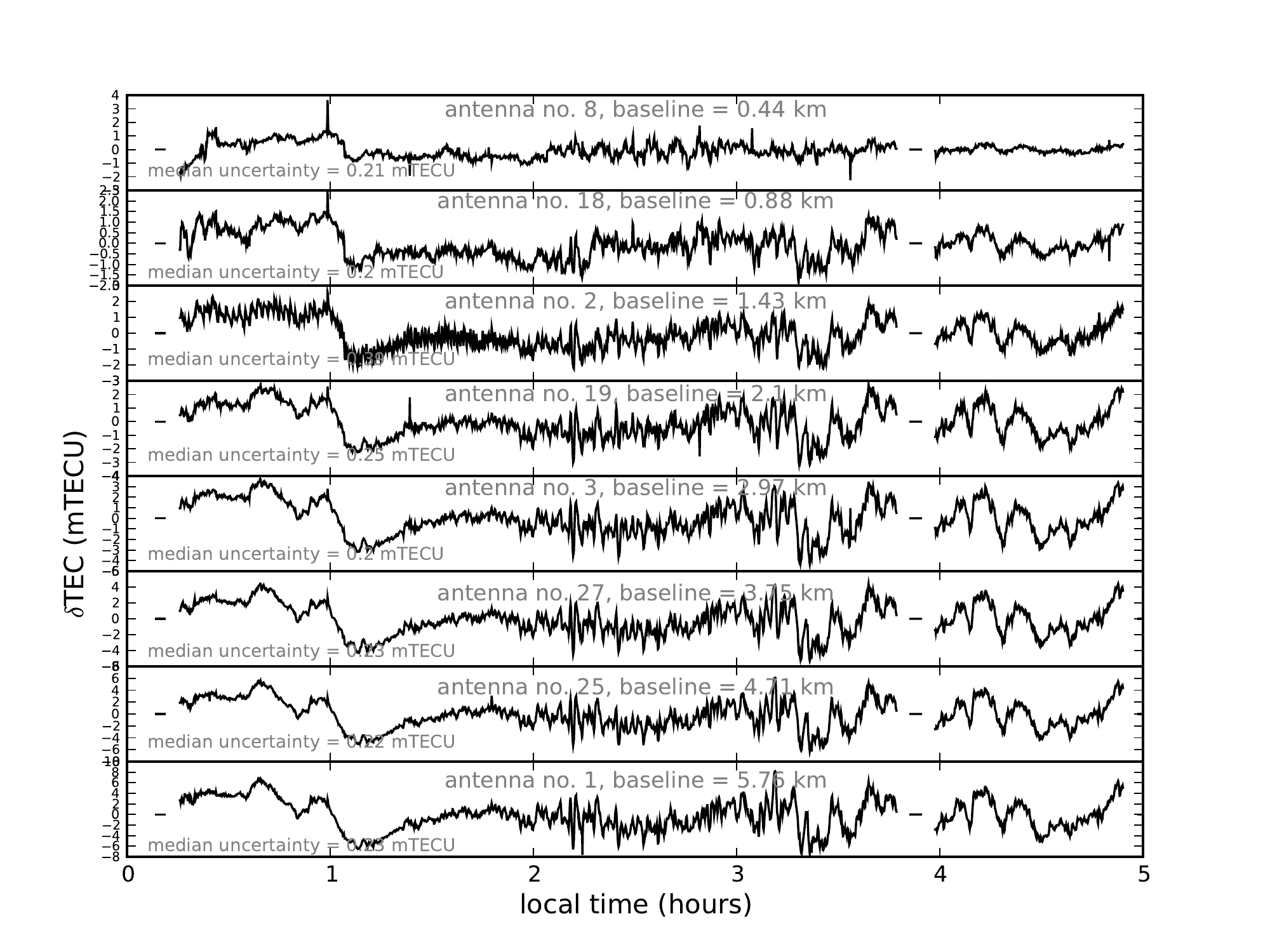}
\caption{For antennas in the northern arm of the VLA, the de-trended values for $\delta \mbox{TEC}$ relative to the reference antenna (see Fig.\ \ref{layout} and Appendix A) as functions of local time.  The baseline lengths and median $1 \sigma$ uncertainties estimated using both polarizations are printed for reference within the panels.}
\label{dtec_n}
\end{figure}

After this, the two polarizations were averaged and divided by a factor of 114 \citep[see, e.g., equation (3) of][]{hel12a} to put them in units of TECU.  The rms deviations between the two polarizations within bins of 10 time steps were used to estimate the $1\sigma$ uncertainties in the $\delta \mbox{TEC}$ values.  Fig.\ \ref{dtec_n} shows examples of $\delta \mbox{TEC}$ time series for antennas in the northern arm of the VLA (see Fig.\ \ref{layout}).  The typical error in the $\delta \mbox{TEC}$ values is about $3\times10^{-4}$ TECU.  This is very similar to what was achieved by \citet{hel12a} using a source that is about ten times brighter than Vir A with similar time averaging (6.67 seconds).  This implies that the  accuracy attained is largely influenced by the time averaging used in the observations and that there are real ionospheric fluctuations on the order of $10^{-4}$ TECU on time scales of about 6.5 seconds.

\subsection{Field-based Calibration}
To obtain some measure of ionospheric fluctuations over the entire $15^{\circ}$ field of view of the 74 MHz VLA, we have also performed a procedure referred to as ``field-based'' calibration.  The principles behind this method and its common application are detailed by \citet{cot04} and \citet{coh07}.  In short, the first order effect of the ionosphere on an image of a cosmic source made from interferometric data is to shift its position.  The observed position shifts of a particular source are  proportional to the components of the TEC gradient along the line of sight to the source averaged over the span of the array and the time interval used to make the image.  At 74 MHz, images of sources with flux densities of more than about 2.5 Jy can typically be made with 1--2 minutes of VLA data to the quality needed to accurately measure the positions of the sources.  Field-based calibration uses a predetermined catalog of sources to choose a set of bright ``calibrators'' and makes images of each calibrator within 1--2 minute intervals, commonly referred to as ``snapshots''.  The position offset from its expected location is then measured for each source and a Zernike polynomial is fit to all of the position offset data for a particular time interval.  The Zernike polynomials are then used to correct the entire field of view for ionospheric effects.\par
For the observation of the field surrounding Vir A, an additional step needed to be performed before running field-based calibration.  Since Vir A is such a bright source, the dominant source of noise within an image made of any source in its vicinity is secondary peaks in the impulse response function (IRF), or ``sidelobes'' from Vir A.  It was therefore necessary to subtract the contribution of Vir A from the observed visibilities before field-based calibration was implemented.  This was done by using the second round of self-calibration to add the effects of the ionosphere to the model visibility data-set computed with the model of the sky brightness distribution of Vir A.  This altered model data-set was then subtracted from the visibility data.  Since the self-calibration process was done in two steps, this subtracted data was still corrected for instrumental effects while the short time-scale ionospheric variations remained.\par
Following the subtraction of Vir A, field-based calibration was performed using the software package Obit (http://www.cv.nrao.edu/$\!\sim\!$bcotton/Obit.html) using one minute snapshots of all calibrators within the field of view brighter than 2.5 Jy at 74 MHz, 48 sources in all (excluding Vir A).  The input source catalog was that produced by \citet{coh07} from the VLA Low Frequency Sky Survey (VLSS) data.  This run of field-based calibration was somewhat compromised by the presence of radio frequency interference (RFI) which is usually seen within 74 MHz VLA data, especially on shorter baselines.  However, a relatively new technique to subtract the effects of stationary sources of RFI from interferometric data has been developed by \citet{ath09} and implemented within Obit.  This technique requires that relatively bright sources be subtracted from the data before it is used, otherwise it may subtract potions of the intensity of such sources at times when their visibility fringe-rates are similar to those of the RFI.  Thus, we used the first run of field-based calibration to subtract out all sources detected within the field of view (this is automatically done within Obit) and ran the RFI subtraction algorithm on the resulting residual data.  We then ran an additional RFI flagging routine within Obit which flags data with unusually large amplitudes, especially those with large Stokes V signals as few if any cosmic VHF sources have non-zero Stokes V intensities.  Following this, the previously subtracted sources were added back to the data, and field-based calibration was performed a second time.  On average, the RFI subtraction and flagging reduced the noise in the calibrator snapshot images by 20\%.\par
\begin{figure}
\noindent\includegraphics[width=\textwidth]{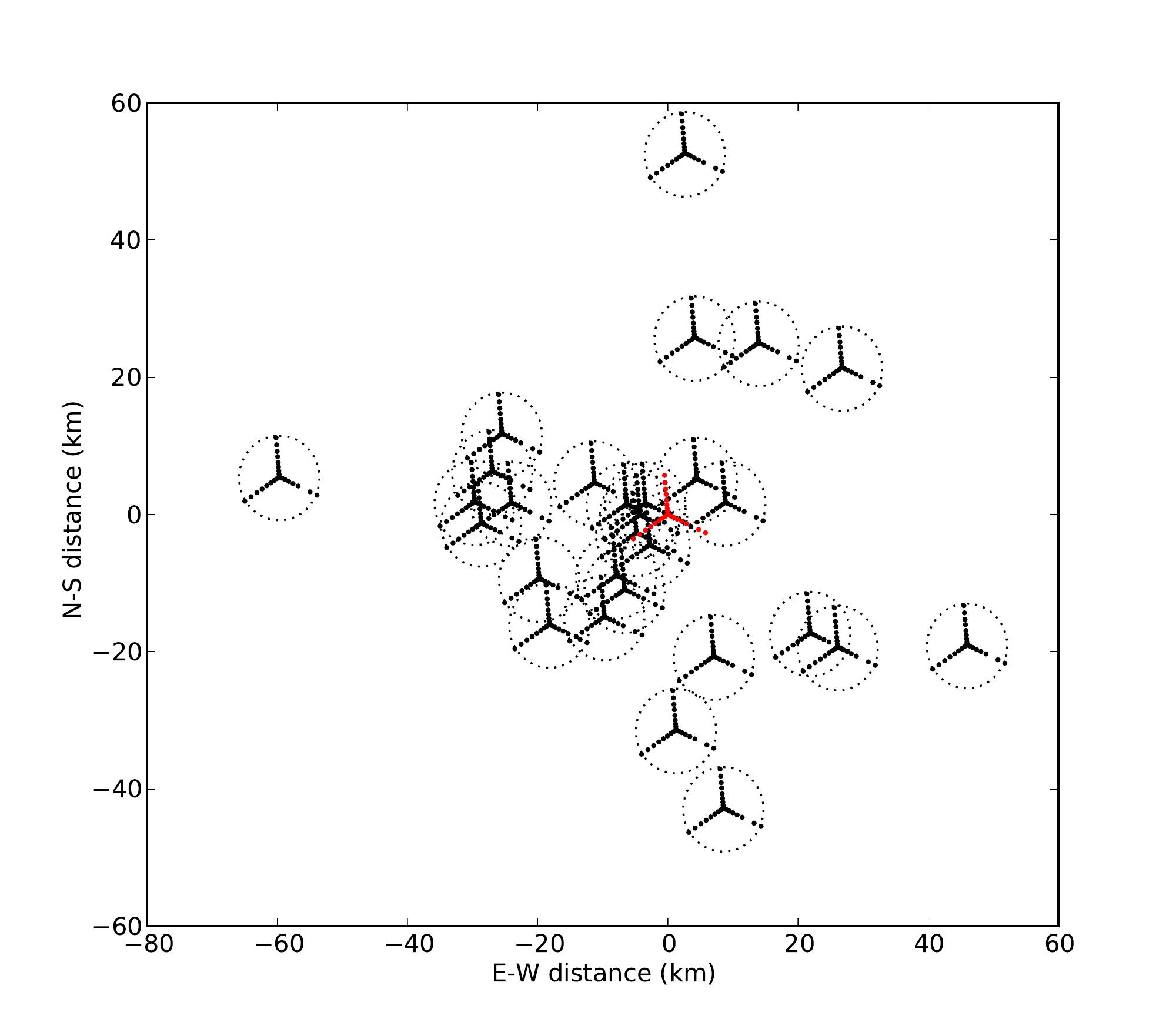}
\caption{The distribution of ionospheric pierce-points at an altitude of 300 km for all 29 calibrator sources used for field-based calibration (see \S 2 and Appendix A) relative to that of Vir A (in red) when it was at its peak elevation in the sky.  The individual VLA antenna positions are plotted relative to each source's pierce-point with a circle drawn to indicate the area covered by the VLA.}
\label{calib}
\end{figure}

Following the completion of field-based calibration, we used the measured calibrator position shifts to compute the TEC gradient as a function of local time for each source.  We limited ourselves to only those sources that were detected in their snapshot images more than half of the time so that the the resulting time series would be reasonably complete which will be important for subsequent spectral analysis (see Appendix B).  There were 29 calibrators that met this criterion.  The relative positions of their associated ionospheric pierce-points are plotted in Fig.\ \ref{calib} for a height of 300 km.  Since we do not know a priori what height most of the observed fluctuations will occur at and the disturbances will likely span a range of heights, the choice of altitude is somewhat arbitrary.  We have chosen 300 km as an approximate location within the lower F region where medium scale disturbances are likely to be found.  In Fig.\ \ref{calib}, the positions of the VLA antennas are also plotted around each calibrator position at a height of 300 km to represent the amount of smoothing that has been effectively applied by using the position offsets to measure the local TEC gradients.\par
After the position shifts were converted TEC gradients assuming a height of 300 km, they were de-trended in a similar manner to the self-calibration-derived $\delta \mbox{TEC}$ values for consistency.  In particular, for each component of the gradient and at each time interval, a line was fit to the data for all time intervals within $\pm$1/4 hour.  Any data with absolute deviations from the fitted line more than three times the median absolute deviation (MAD) were then identified and the fit was redone without them.  If the absolute difference between the value at the current time interval and the final linear fit was $<\!5\times$MAD, the fitted value was subtracted from it.  Otherwise, this time interval was flagged as being possibly spurious.\par
\begin{figure}
\noindent\includegraphics[width=\textwidth]{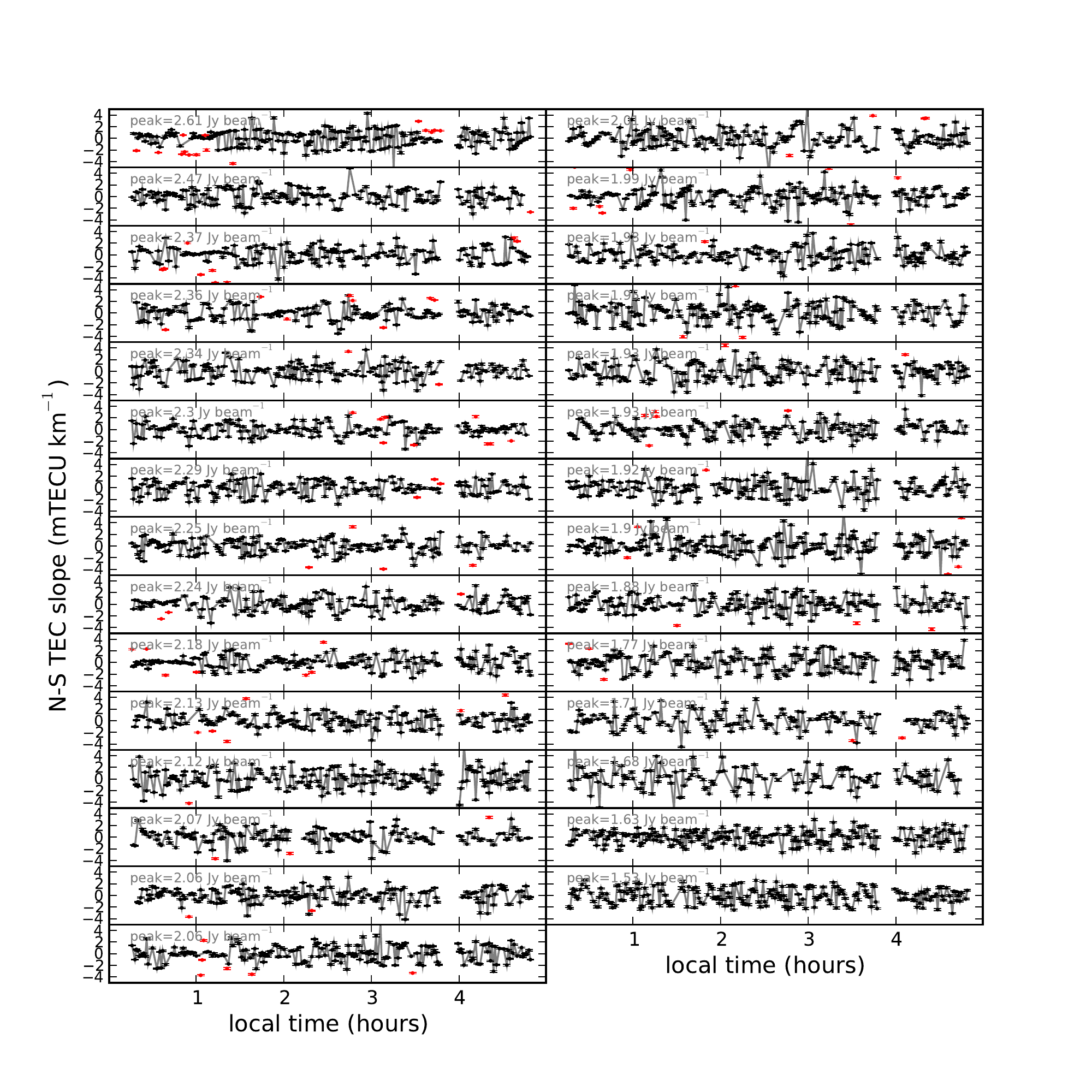}
\caption{For the 29 calibration sources, the de-trended and flagged (red points) north-south components of the TEC gradient (north is positive) as functions of local time.  The peak intensity of each source is given in each panel in units of Jansky (1 Jy = $10^{-26}$ W m$^{-2}$ Hz$^{-1}$).}
\label{grad_ns}
\end{figure}

As an example of the results of this processing, we have plotted the de-trended north-south components of the TEC gradients as functions of time for all 29 calibrators in Fig.\ \ref{grad_ns} with the source peak intensities printed in each panel.  Note that these are the observed peak intensities which are lower than the sources' cataloged intensities because of the response of the individual antennas which decreases significantly with angular separation from the center of the field of view.  Flagged time intervals are highlighted in red.  These data include error-bars computed using the errors in the position offsets.  These were computed at each time interval according to \citet{con97}, where the error in the source position is given by
\begin{equation}
\sigma_p \approx \frac{1}{2} \sigma \theta / S_p
\end{equation}
where $\sigma$ is the noise in the image, $\theta$ is the angular diameter of the source, and $S_p$ is the peak intensity on the image.  Here, we have assumed that the sources are unresolved and consequently have the same angular diameter as the IRF, in this case, 76 arcseconds.  The dominant source of noise within each snapshot is sidelobes from other sources within the filed of view.  To estimate the typical value of $\sigma$ for the snapshots, we made an image of the full filed of view using all 270 minutes of data with no deconvolution applied (i.e., without removing source sidelobes).  The rms noise for this image was 0.042 Jy beam$^{-1}$, implying $\sigma\approx0.7$ Jy beam$^{-1}$ for each one-minute snapshot.  Given the intensities of the calibrators used (see Fig.\ \ref{grad_ns}), this further implies a typical TEC gradient uncertainty of about $10^{-4}$ TECU km$^{-1}$.

\section{Analysis Methods}
We have performed separate spectral analyses on both the self-calibration and field-based calibration data.  As we will detail below, the methods used are quite complementary and we include some comparison of the two in the main body of this article.
\subsection{Self-calibration Data}
The spectral analysis of the de-trended $\delta \mbox{TEC}$ data derived from self-calibration using Vir A (see Fig.\ \ref{dtec_n} for examples) was performed using two complementary techniques.  The first follows that developed by \citet{hel12b} and the second uses a straightforward Fourier-style analysis.  Both techniques start with a second order, two-dimensional polynomial fit over the array to obtain approximate values for the TEC gradient at each antenna's associated pierce-point.  The locations of these pierce-points were computed assuming a nominal height of 300 km.  However, since the lines of sight from the antennas to the sources are essentially parallel, the effect of the assumed height on these computed positions is marginal unless the observed source is at a relatively low elevation.  During the VLA observations, Vir A was no fewer than $45^\circ$ above the horizon, indicating that the projected antenna separations changed very little throughout the observations \citet[see][]{hel12a} and had only a weak dependence on assumed height.  For instance, the relative pierce-point positions among the antennas for a height of 300 km differ by at most a few percent from those computed for a height of 3000 km.  The time series for the coefficients of the polynomials fit using these pierce-point positions were then Fourier transformed (here, within a 1/2 hour sliding window) to obtain a complex spectrum of the gradient for each antenna.
\subsubsection{Phase-based Technique}
The two spectral analysis techniques both use the temporal Fourier transforms of the antenna-based TEC gradient time series.  The first technique from \citet{hel12b} uses the phase information from the Fourier transform at each time step to estimate the weighted mean direction and magnitude of the wavenumber vector, $\vec{k}$, for each temporal mode.  These results were combined with the temporal frequencies to also estimate pattern speeds.  This technique uses both components (i.e., north-south and east-west) components of the TEC gradient.  The technique assumes that the line along which the phases of the Fourier transforms of the two components agrees is the direction for that temporal mode.  This is done by first fitting a plane to the Fourier transform phase of each TEC gradient component after unwrapping the phases along each VLA arm.  Then, the parameters of the two planar fits are used to determine the line defined by the intersection of the two planes.\par
It was shown by \citet{hel12b} that this phase-based approach can mitigate the influence of wavefront distortions on the analysis of relatively large waves (i.e., those substantially larger than the array).  However, it also has the potential to sense the direction of motion for non-wavelike patterns.  Imagine, for instance, a distortion of arbitrary shape passing over the array.  If one chooses a series of points along its direction of motion, the time series will be identical for each point except that they will be out of phase with one another.  If the velocity is constant, then the phase difference between two of these points will just be proportional to the separation between them.  This holds true for the TEC times series as well as for the time series of the two components of the gradient.  Therefore, the ``phase-based'' technique of \citet{hel12b} is capable of estimating the direction of motion, rather than orientation, when non-wavelike disturbances are encountered.
\subsubsection{DFT-based Technique}
The second technique uses a two-dimensional, spatial discrete Fourier transform (DFT) of each of the TEC gradient components measured at each antenna\footnote{Formally, an inverse DFT was used since a particular oscillating mode $\sim\!\mbox{exp}[2\pi i (\nu t - x\xi_x - y\xi_y)]$}.  Since the array is ``Y''-shaped, one can imagine that the IRF for this DFT is rather poorly behaved.  We show in Fig.\ \ref{dft} that this is in fact the case.  The left panel of Fig.\ \ref{dft} displays the IRF for the first half-hour of VLA data.  As one might expect, there are substantial sidelobes in the regions between VLA arms extending to relatively large spatial frequencies.  To mitigate the effect of these sidelobes, we have implemented a version of the clean algorithm \citep{cbb99}.  This algorithm is often used with images produced using radio interferometers to remove the effects of sidelobes associated with bright sources.  Briefly, it is an iterative algorithm that identifies a peak in the image plane at each iteration and subtracts a scaled version of the IRF centered on that peak where the scale factor is usually the intensity multiplied by a predetermined gain value.  The iterations are continued until a maximum iteration limit is reached or some convergence criteria are met.  Often, this is done interactively using boxes drawn by eye to limit the locations of the model delta functions, or ``clean components'' to specific regions.  After the cleaning is finished, the clean components are convolved with a more well-behaved IRF, usually a Gaussian whose parameters are determined from the main peak of the IRF.  Following this, the re-convolved components are added back to the image.\par
Our implementation is only slightly different since the spatial spectra we are ``cleaning'' are complex and not real valued functions like the images made with interferometers.  Consequently, at each iteration, we used the spectral power to determine the location of the next clean component to be subtracted, and the clean components all had complex amplitudes.  For each clean component, a scaled version of the IRF was used that was computed using the same grid spacing as the spectra but with $2N+1$ pixels where $N$ is the number used for the actual spectra (here, $N=256$; see Fig.\ \ref{dft}).  This was done so that the IRF could be easily shifted to be centered directly on a particular spectral pixel.  After trying several convergence criteria, maximum iteration limits, and gain values, we found that satisfactory results were obtained by simply running 20 clean iterations with a gain of 0.5 and no limits on the locations of the clean components (i.e., no ``boxing'').  This cleaning was run separately on each temporal frequency and TEC gradient component within each time bin.  After restoring the re-convolved clean components to their respective spectra, the power spectra from both components of the TEC gradient were combined into a single gradient power spectrum for each temporal frequency and time bin.\par
Fig.\ \ref{dft} shows examples of the results of the clean algorithm applied to the first half-hour of self-calibration data.  In the right panels, we show the spectral power for the strongest temporal frequency, 6.375 hr$^{-1}$, and for a randomly chosen frequency, 75.374 hr$^{-1}$, both before and after cleaning.  From these, on can see that nearly all the structure seen within the raw spectral maps was from the IRF.  It also shows that the overarching assumption of \citet{hel12b} that each temporal mode is well approximated by a single wavenumber vector is essentially correct.  Because of this, and so we can more directly compare the two methods, we used the cleaned spectral power to computed the peak power and a single weighted (with the spectral power) mean value of the wavenumber vector for each temporal mode for each time step.  Because it is a more conventional Fourier-based analysis, this ``DFT-based'' method is most sensitive to wavelike structures and can only measure the wavefront orientation, which dow not necessarily give the direction of motion.  The importance of this distinction is made apparent in the discussion in \S 3.\par
Because both techniques estimate a single wavenumber vector for each temporal mode, they essentially use both the apparent motion of Vir A and the motion of the observed disturbances to convert the properties of temporal oscillations observed with a compact array into spatial information.  This allows a much higher spatial frequency resolution to be achieved than could normally be obtained with such a small array.  For instance, the temporal spectral resolution, $\Delta \nu$, is related to its spatial counterpart, $\Delta \xi$, simply via the pattern speed, i.e., $\Delta \xi = \Delta \nu / v$.  For the 1/2 hour wide window we have used here, $\Delta \nu \approx 2 \mbox{ hr}^{-1}$.  Therefore, for a disturbance traveling 100 m s$^{-1}$, $\Delta \xi \approx 0.0055 \mbox{ km}^{-1}$ which is the equivalent of what can be obtained with an array with a diameter of nearly 200 km.  The results from both techniques are discussed in more detail in \S 2.
\subsection{Field-based Calibration Data}
While the self-calibration-based spectral analysis supplies results with remarkable sensitivity and spatial frequency resolution, it suffers from two significant drawbacks.  First, even though it uses a relatively long temporal baseline to achieve the equivalent of a long spatial baseline, the measurements are still confined to a narrow strip, in this case, about 11 km wide.  This makes it difficult to accurately estimate the direction and speed of relatively large disturbances in the presence of significant wavefront distortions.  Second, wave properties are derived from the data for each temporal frequency.  This implies that if there are several wave patterns with different sizes but similar periods, there is no way within this method to separate them from one another and all that can be estimated is a combined speed and direction for all of them.  For isotropic phenomena such as turbulence, this may be acceptable, but for characterizing wave properties, it can be problematic, especially at temporal frequencies where there is no clearly dominant wave pattern.\par
What is needed to supplement this analysis is another array that provides TEC gradient measurements over similar time intervals but spread out over a much larger area.  The position offset data used in the field-based calibration method detailed in Appendix A provides just such an array.  As Fig.\ \ref{calib} demonstrates, the distribution of calibrator sources on the sky provides and array of TEC gradient measurements that spans an area with a diameter of about 100 km or more (at a height of 300 km) with a quasi-random distribution of ionospheric pierce-points.  With this data, it is therefore possible to perform a relatively straightforward Fourier analysis on the data without utilizing any ``tricks'' to retrieve spatial information about the observed TEC fluctuations.  One may simply perform a Fourier transform in three dimensions, one temporal and two spatial, over a given time interval.  This will yield spectral ``cubes'' so that different spatial modes with similar temporal frequencies may be separated from one another.  We note that the field-based data has effectively been smoothed in time (with a one minute window) and space (with an 11 km diameter circular window) so that higher frequency modes are damped out.  However, the self-calibration data provide ample information about such modes, making the two analyses quite complementary.

\subsubsection{Method}
To begin the analysis of the field-based data, we first put the data into 8 temporal bins.  To minimize sidelobes in the temporal components of the Fourier transforms, we have weighted the TEC gradient time series within each bin by a unit amplitude Gaussian with a full width at half maximum (FWHM) of 1/2 hour.  We therefore constructed the temporal bins to be one hour wide, but separated by 1/2 hour so that all time intervals would have Gaussian weights $>\!0.5$ in at least one time bin.  For each source, there were also time intervals where either the source's position could not be measured, or the data was otherwise flagged as spurious (see Appendix A).  Such gaps in temporal coverage can also lead to increased sidelobes in frequency space.  We therefore reduced the Gaussian weight of any time interval adjacent to such a gap by a factor of two to introduce some amount of tapering near these gaps to mitigate their effects on the impulse response.  Using this weighting scheme, we performed a discrete Fourier transform (DFT) of each component of the TEC gradient time series for each source within each of the 8 time bins at 60 frequencies up to the Nyquist limit of 30 hr$^{-1}$.  In each instance, we also computed the amplitude of the impulse response at $\nu = 0$ and normalized each complex temporal spectrum by this value.\par
\begin{figure}
\noindent\includegraphics[width=\textwidth]{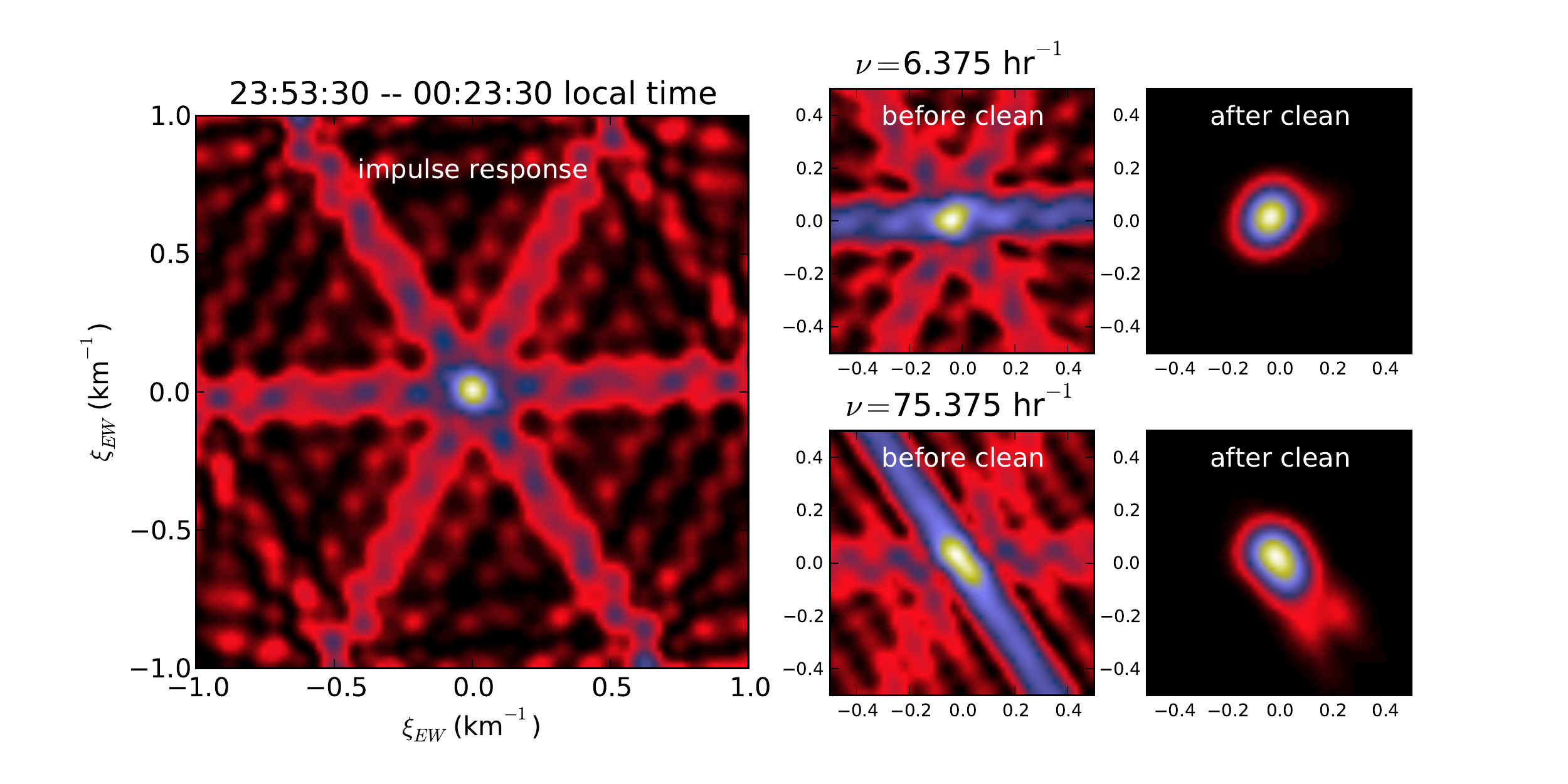}
\caption{An example of the DFT-based spectral analysis on the first half-hour of self-calibration data.  The left panel shows the impulse response from the DFT of the antenna positions.  The panels to the right show the spectral power as functions of north-south and east-west spatial frequency for the brightest temporal frequency, 6.375 hr$^{-1}$, and one chosen at random, 75.375 hr$^{-1}$, both before and after the clean algorithm was applied.}
\label{dft}
\end{figure}
\begin{figure}
\noindent\includegraphics[width=\textwidth]{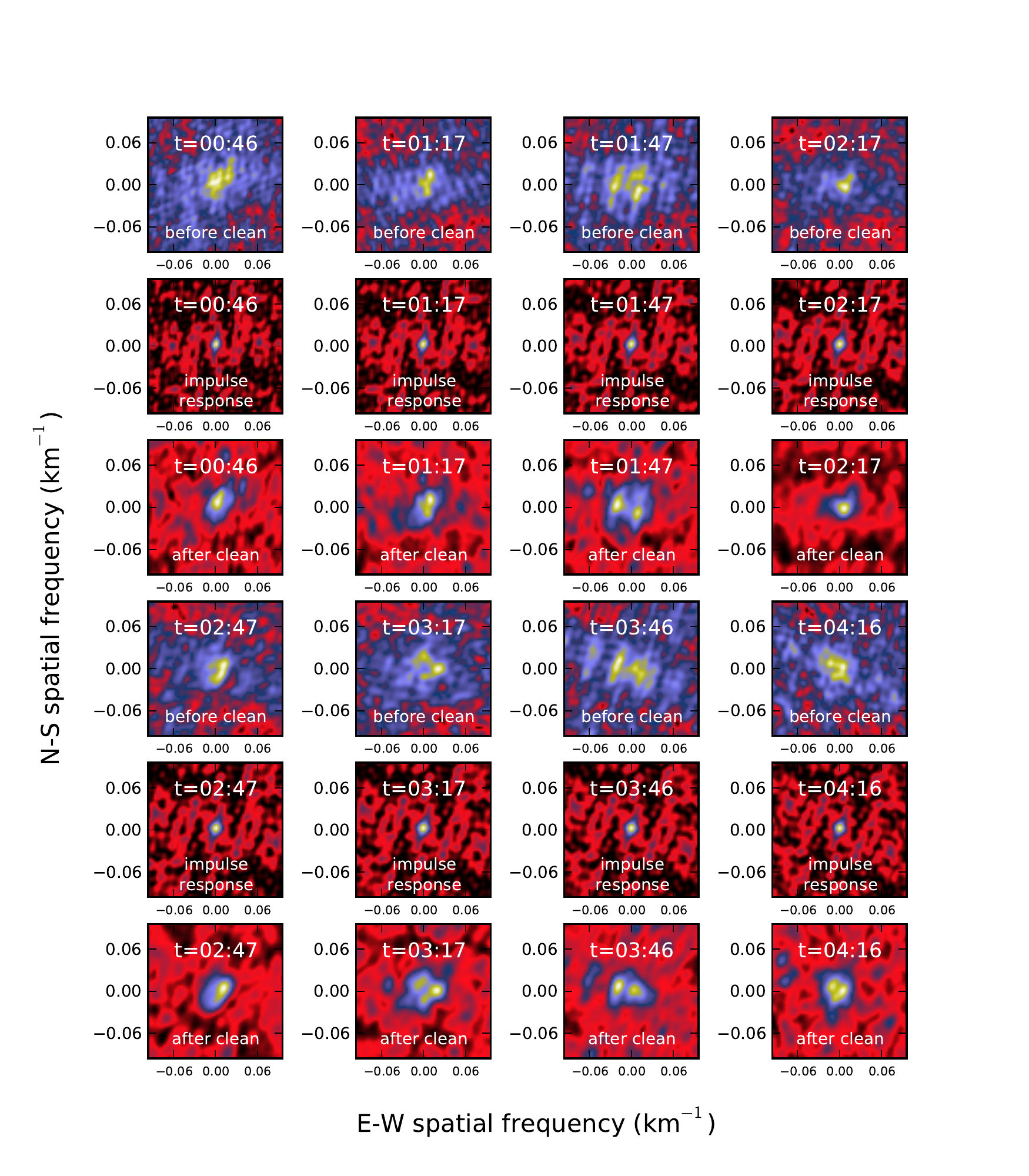}
\caption{Within 1/2 hour windows, the mean (over all temporal frequencies) TEC gradient power spectra determined from the field-based calibration data (see Appendix A and Fig.\ \ref{grad_ns} for examples) before and after the application of the clean de-convolution algorithm (see Appendix B).  The IRF for each spectrum is also displayed for reference.}
\label{clean}
\end{figure}

Following the generation of temporal spectra for the calibrators, the mean north-south and east-west position of each calibrator's pierce-point relative to that of the center of the filed of view at a height of 300 km was computed for each time bin.  These were then used with a DFT to compute a complex spectrum on a grid of north-south and east-west spatial frequencies for each time bin and temporal frequency and for each component of the TEC gradient.  Again, these spectra were normalized by the amplitude of the IRF at a frequency of zero.  Given the relatively sparse sampling provided by the distribution of calibrators, each resulting spatial spectrum was significantly affected by sidelobes.  We sought to mitigate these effects using an implementation of the clean algorithm as was done with the DFT-based method that was applied to the self-calibration data (see \S B1 and Fig.\ \ref{dft}).  As before, 20 iterations were used per temporal frequency and gradient component with a gain of 0.5.  In Fig.\ \ref{clean}, we show the mean gradient power over all temporal frequencies at each time bin both before and after cleaning.  For reference, we have also included images of the IRF power spectra.  One can see that the cleaning process has substantially improved the fidelity of the spectral images.\par
\begin{figure}
\noindent\includegraphics[width=\textwidth]{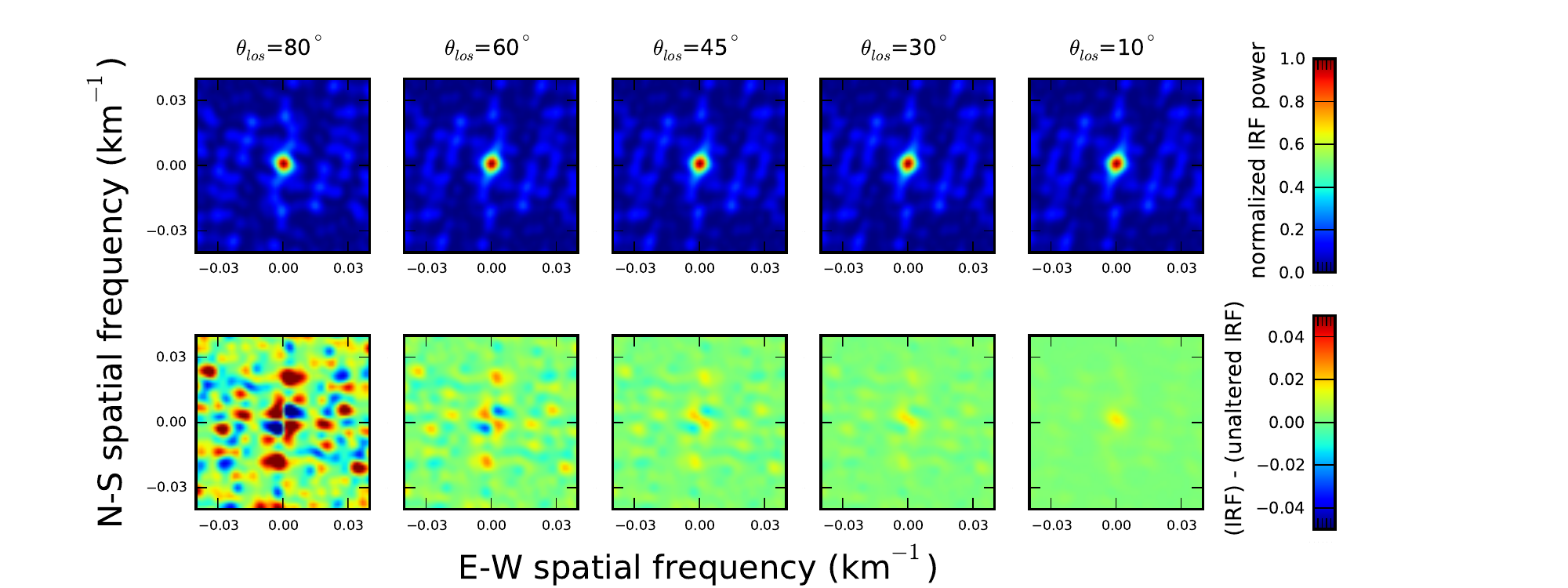}
\caption{Examples of the impact of different viewing angles of plane wave-like disturbances on the measured impulse response for the spectral analysis of field-based data.  The upper panels show the measured IRF for different values of the angle between the line of sight to the center of the field of view and the optimum viewing angle for a given disturbance, $\theta_{los}$.  The bottom panels show the difference between this IRF and the unaltered version (see \S B2.1 for more details).}
\label{orien}
\end{figure}

Unlike the self-calibration data, the multiple lines of sight used in this analysis are far from parallel.  This means that each of them ``views'' any disturbance from a different angle causing the observed amplitude of such a disturbance to vary across the VLA field of view.  For instance, if there is an optimum viewing angle for a particular disturbance, the detected amplitude for a given line of sight is $\propto\! \mbox{cos}(\theta_{los})$, where $\theta_{los}$ is the angle between the actual line of sight and the optimum one (i.e., the dot product between the unit vectors for the actual and optimum lines of sight).  The effect of this on our analysis is that this position dependent weighting of the measured gradients alters the IRF from its expected shape.  To illustrate the magnitude of this effect, we show in the panels of Fig.\ \ref{orien} the observed IRF for values of $\theta_{los}$ ranging from $10^\circ$--$80^\circ$.  For this demonstration, we assumed the optimum view angle was toward the north (i.e., its unit vector had no east-west component).  As expected, the most notable change is for the largest $\theta_{los}=80^\circ$.  The lower panels of Fig.\ \ref{orien} show this better where they display the difference between the observed IRF (scaled to a maximum of unity) and the expected IRF.  The maximum difference for $\theta_{los}=80^\circ$ is nearly 10\%, but ranges from $<\!\!1$\% to a few percent for the other values of $\theta_{los}$.\par
Therefore, when we clean the field-base spectral cubes, we are basically using the wrong IRF, leading to sidelobes that are not properly subtracted and even adding additional but weak sidelobes.  Since there are typically several detected features within each field-based spectral cube which likely have a distribution of optimum viewing angles, the net effect of this on our analysis is to increase the noise of the spectral maps.  However, as Fig.\ \ref{orien} indicates, this is likely a relatively small effect which is typically on the order of 1\% and only as high as 10\% in extreme cases.

\begin{acknowledgments}
Basic research in astronomy at the Naval Research Laboratory is supported by 6.1 base funding.  The National Radio Astronomy Observatory is a facility of the National Science Foundation operated under cooperative agreement by Associated Universities, Inc.  The authors are grateful to MIT Haystack Observatory for GPS TEC data obtained from their Madrigal data center.
\end{acknowledgments}

\end{article}

\clearpage

\clearpage

\clearpage

\clearpage

\clearpage

\clearpage

\clearpage

\clearpage

\clearpage

\clearpage

\clearpage

\clearpage

\clearpage

\clearpage

\clearpage

\end{document}